\documentclass[%
 reprint,
superscriptaddress,
 amsmath,amssymb,
 aps,
]{revtex4-2}

\usepackage{graphicx}
\usepackage{dcolumn}
\usepackage{bm}
\usepackage{braket}
\usepackage{color}
\usepackage{amsthm}
\usepackage{dsfont}
\usepackage[absolute]{textpos}
\usepackage{xcolor}
\usepackage{fancyhdr}

\DeclareMathOperator{\Tr}{Tr}

\def\bx{\mathbf{x}}

\def\bw{\mathbf{w}}
\def\bu{\mathbf{u}}
\def\bv{\mathbf{v}}
\def\balpha{\bm{\alpha}}

\def\bbD{{\bm{\mathcal{D}}}}

\usepackage{appendix}

\begin{document}

\preprint{APS/123-QED}

\title{Improving application performance with biased distributions of quantum states}

\author{Sanjaya Lohani}
\thanks{These two authors contributed equally}
\affiliation{IBM-HBCU Quantum Center, Howard University, Washington, DC 20059, USA}
\affiliation{Tulane University, New Orleans, LA 70118, USA}
\author{Joseph M. Lukens}
\thanks{These two authors contributed equally}
\affiliation{Quantum Information Science Group, Oak Ridge National Laboratory, Oak Ridge, Tennessee 37831, USA}
\author{Daniel~E. Jones}
\affiliation{United States Army Research Laboratory, Adelphi, MD 20783, USA}
\author{Thomas~A. Searles}
\affiliation{IBM-HBCU Quantum Center, Howard University, Washington, DC 20059, USA}
\affiliation{Massachusetts Institute of Technology, Cambridge, MA 02139, USA}
\author{Ryan~T. Glasser}
\email[]{rglasser@tulane.edu}
\affiliation{Tulane University, New Orleans, LA 70118, USA}
\author{Brian~T. Kirby}
\email[]{brian.t.kirby4.civ@mail.mil}
\affiliation{Tulane University, New Orleans, LA 70118, USA}
\affiliation{United States Army Research Laboratory, Adelphi, MD 20783, USA}

\date{\today}

\begin{abstract}
We consider the properties of a specific distribution of mixed quantum states of arbitrary dimension that can be biased towards a specific mean purity. 
In particular, we analyze mixtures of Haar-random pure states with Dirichlet-distributed coefficients. We analytically derive the concentration parameters required to match the mean purity of the Bures and Hilbert--Schmidt distributions in any dimension.
Numerical simulations suggest that this value recovers the Hilbert--Schmidt distribution exactly, offering an alternative and intuitive physical interpretation for ensembles of Hilbert--Schmidt-distributed random quantum states.
We then demonstrate how substituting these Dirichlet-weighted Haar mixtures in place of the Bures and Hilbert--Schmidt distributions results in measurable performance advantages in machine-learning-based quantum state tomography systems and Bayesian quantum state reconstruction.
Finally, we experimentally characterize the distribution of quantum states generated by both a cloud-accessed IBM quantum computer and an in-house source of polarization-entangled photons.  
In each case, our method can more closely match the underlying distribution than either Bures or Hilbert--Schmidt distributed states for various experimental conditions.
\end{abstract}

\maketitle
\begin{textblock}{13.3}(1.4,15)
\noindent\fontsize{7}{7}\selectfont \textcolor{black!30}{This manuscript has been co-authored by UT-Battelle, LLC, under contract DE-AC05-00OR22725 with the US Department of Energy (DOE). The US government retains and the publisher, by accepting the article for publication, acknowledges that the US government retains a nonexclusive, paid-up, irrevocable, worldwide license to publish or reproduce the published form of this manuscript, or allow others to do so, for US government purposes. DOE will provide public access to these results of federally sponsored research in accordance with the DOE Public Access Plan (http://energy.gov/downloads/doe-public-access-plan).}
\end{textblock}

\section{Introduction}

Ensembles of random density matrices have found broad applicability in quantum information science \cite{montanaro2007distinguishability,hamma2012quantum,miatto2015recovering,girolami2011quantum,kirby2016entanglement,lu2011optimal,roncaglia2014bipartite}.  Of particular recent interest is their use in quantum state tomography systems, either as training sets for machine-learning-based techniques \cite{lu2018separability,lohani2020machine,danaci2021machine,lohani2020experimental,ahmed2020classification} or as prior distributions for Bayesian state reconstruction \cite{Blume2010, Seah2015,Granade2016,Williams2017,mai2017pseudo,lukens2020practical, Lu2020b}.  Significant effort has been devoted to developing such ensembles based on various underlying measures and determining their characteristics \cite{wootters1990random,zyczkowski2001induced,Sommers2003,zyczkowski2005average,al2010random}.  However, as quantum information science becomes increasingly reliant on classical computational resources for support, the opportunity has emerged to improve the performance of these systems through the creation of bespoke ensembles that more closely resemble the system under investigation. 

In the case of pure states, ensembles are typically generated according to the Fubini--Study measure, induced by the Haar measure over the unitary group $U(D)$ \cite{wootters1990random,zyczkowski2001induced,al2010random}. A random pure state $\vert \psi_{R}\rangle$ can be found by first generating a Haar-random unitary from $U(D)$ and applying it to any fixed quantum state $\vert\psi\rangle$ to obtain $\vert\psi_{R}\rangle=U\vert\psi\rangle$.
Equivalently, one could use a column of the Haar-distributed unitary as $\vert\psi_{R}\rangle$, since $\vert\psi\rangle$ can be chosen arbitrarily. 

Several methods for generating ensembles of mixed states exist.
Two of the most widely used are based on the Hilbert--Schmidt (HS) and Bures measures.
The HS measure has a simple physical interpretation as that which is induced when a Haar-random pure state $\vert\psi\rangle$ of dimension $D^{2}$ is traced down to dimension $D$.
More generally, the HS measure is a special case of a family of induced measures related to tracing a $DM$-dimensional Haar-random pure state down to a $D$-dimensional mixed state.
Although somewhat less intuitive physically, another metric commonly used for generating ensembles of random density matrices is the Bures measure.
The Bures measure is unique as the sole monotone metric that is both Fisher and Fubini--Study adjusted, thereby aligning with standard metrics in both the classical and pure-state limits~\cite{Sommers2003}. In this sense, the Bures distribution represents a canonical choice for situations of complete randomness, such as a prior for Bayesian state reconstruction when all possible input states can occur.  

Recently, an additional ensemble of mixed quantum states based on sums of nonorthogonal Haar-random pure states 
has been explored as a prior for Bayesian quantum state reconstruction techniques, 
where the coefficients of these ensembles are distributed according to the symmetric Dirichlet distribution~\cite{mai2017pseudo,lukens2020practical,Lu2020b}. Motivated initially by both its computational simplicity and amenability to tuning effective rank~\cite{mai2017pseudo}, this distribution allows for significant speed-ups in Bayesian quantum state estimation compared to alternative parameterizations~\cite{lukens2020practical}.
Yet it is not clear from the existing literature how these states behave outside of limiting special cases, particularly in relation to other standard distributions, thus leaving a significant gap in the theoretical justification for this distribution.

In this paper, we determine various properties of the ensembles of density matrices generated by mixing Haar-random pure states with Dirichlet-distributed coefficients.  
We find that an immediate advantage of these ensembles compared to standard alternatives is that they can be biased toward a particular mean purity.
We also determine the Dirichlet concentration parameter required to recover mean purities equal to those of the standard Bures or HS measures.
Numerical simulation of up to ten-dimensional states indicates that for proper tuning of the concentration parameter, we can exactly recover the HS distribution, offering an alternative and intuitive physical interpretation for random quantum states distributed according to the HS measure.
We then apply our results to generate training data for a machine-learning-based quantum state tomography system and as a prior distribution for Bayesian state reconstruction.  
In each case, we find measurable performance improvements using tailored ensembles of states as opposed to those based on Bures or HS.  
Finally, we perform experiments on both an IBM quantum computer (IBM Q) and a source of polarization-entangled photons to determine the typical ensemble of states generated by each throughout a characteristic experiment.  
In each case, we find that we can more closely match the distributions of these systems using Dirichlet-distributed mixtures than by standard methods.  

\section{Biased distributions}

\subsection{Dirichlet distribution}

The Dirichlet distribution is defined for vectors $\bx=(x_{1},...,x_{K})$ where the elements of $\bx$ belong to the open $K-1$ simplex: $x_j\geq 0$ and  $\sum_{j=1}^K x_j = 1$. 
The probability density function of the Dirichlet distribution is defined by
\begin{equation}
    \text{Dir}(\bx\vert\balpha)=\frac{\Gamma\left(\sum_{j=1}^{K}\alpha_{j}\right)}{\prod_{j=1}^{K}\Gamma(\alpha_{j})}\prod_{j=1}^{K}x_{j}^{\alpha_{j}-1},
    \label{eq:dirichlet_def}
\end{equation}
where $\balpha=(\alpha_{1},...,\alpha_{K})$ with all $\alpha_{j}\ge 0$ defines the concentration parameters and $\Gamma(\cdot)$ is the standard gamma function.

\begin{figure}[!tb]
\centering 
	\includegraphics[width=\columnwidth]{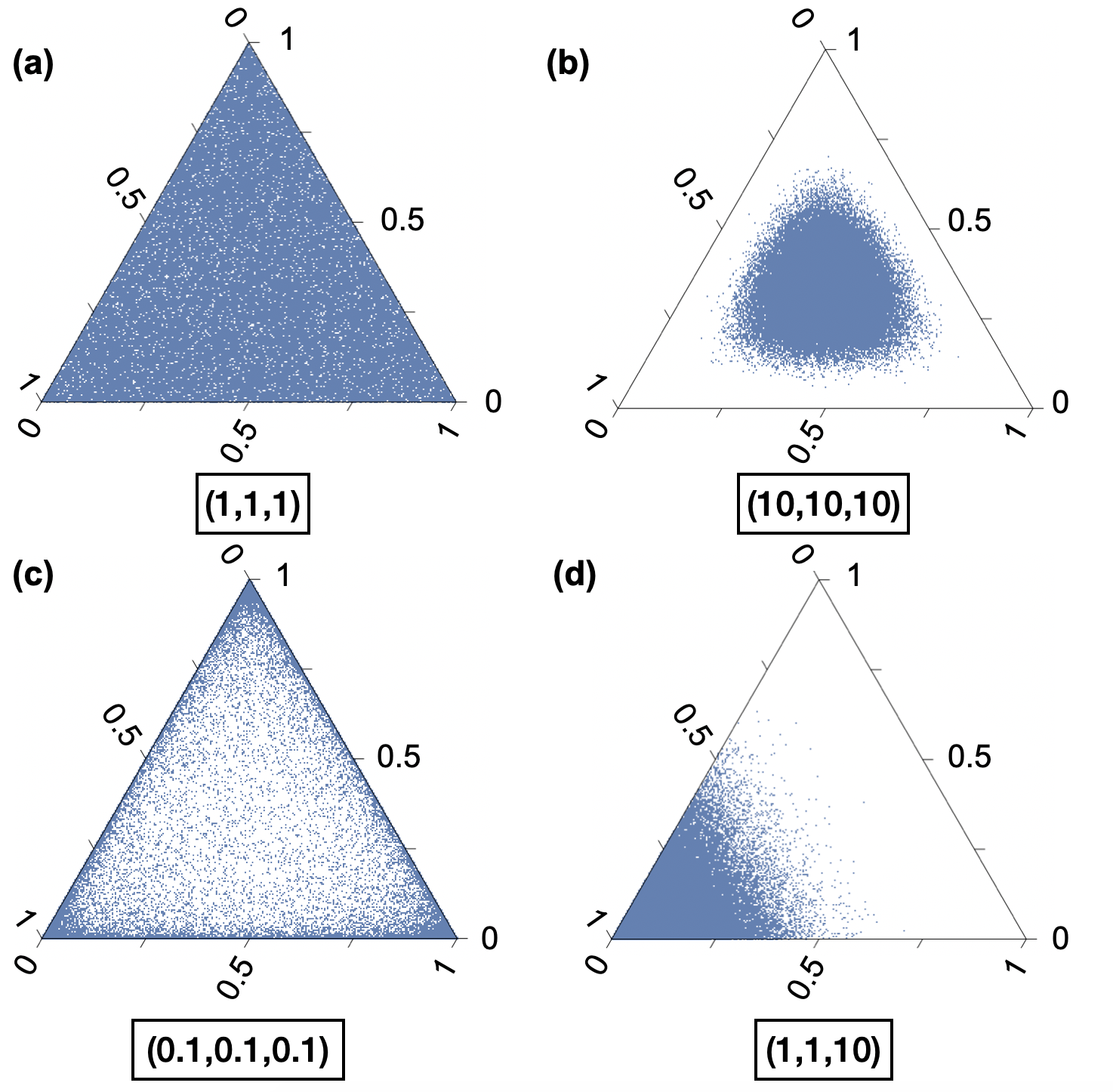}
	\caption{Ternary plots of $10^5$ Dirichlet vectors $\bx=(x_{1},x_{2},x_{3})$.  Each plot is sampled from a distribution with a different set of concentration parameters, with (a)--(c) symmetric.}
	\label{Fig:Dirichlet_Quad}
\end{figure}

The concentration parameters determine the properties of the distribution. 
To understand this intuitively, in Fig.~\ref{Fig:Dirichlet_Quad} we plot $10^5$ Dirichlet-distributed random points on a ternary plot when $K=3$. 
The four examples shown in Fig. \ref{Fig:Dirichlet_Quad} are representative exceptional cases. 
In Fig. \ref{Fig:Dirichlet_Quad}(a), the distribution is uniform, with an equal probability of sampling anywhere within the simplex.
The uniform sampling of Fig.~\ref{Fig:Dirichlet_Quad}(a) contrasts with Fig.~\ref{Fig:Dirichlet_Quad}(b--d), where the concentration parameters bias the distribution towards the center, away from the center, or into one corner of the simplex, respectively.   

The behavior visualized in Fig.~\ref{Fig:Dirichlet_Quad} is also reflected in the mean, variance, and covariance of the Dirichlet distribution.
These statistical properties will be useful in what follows and are given by
\begin{equation}
\begin{aligned}
    \text{E}\left[X_{j}\right]&=\tilde{\alpha}_{j}\\
    \text{Var}\left[X_{j}\right]&=\frac{\tilde{\alpha}_{j}(1-\tilde{\alpha}_{j})}{\alpha_{0}+1}\\
    \text{Cov}\left[X_{j},X_{k}\right]&=\frac{\delta_{jk}\tilde{\alpha}_{j}-\tilde{\alpha}_{j}\tilde{\alpha}_{k}}{\alpha_{0}+1}
\end{aligned}
\label{eq:stat}
\end{equation}
for all $j,k\in\{1,...,K\}$, with $\delta_{jk}$ the Kronecker delta, $\alpha_{0}=\sum_{j=1}^{K}\alpha_{j}$, and $\tilde{\alpha}_{j}=\alpha_{j}/\alpha_{0}$.
In the equations below, we note that we do not always use capital letters to distinguish random variables from the specific values they assume when the result is more clear otherwise.

Throughout this paper, we will be interested in special cases of the Dirichlet distribution with all $\alpha_j$ equal, known as the symmetric Dirichlet distribution. 
Figure~\ref{Fig:Dirichlet_Quad}(a--c) are examples of the symmetric Dirichlet distribution, which behave analogously for any value of $K$.
In particular, the flat Dirichlet distribution occurs when all concentration parameters are equal to $1$ for any $K$.
Although extensions of our results beyond the symmetric distribution are straightforward, we restrict to the symmetric distribution in what follows because the equivalence under exchange of index simplifies the presentation.  

\subsection{Constructing mixtures of Dirichlet-weighted pure states}

The properties of the Dirichlet distribution described above make it remarkably well suited for defining the coefficients of a mixture of pure quantum states.
To begin with, the restriction of the components of $\bx$ to the $K-1$ simplex is precisely the requirement for the coefficients of a convex sum, such as a density matrix which is the convex sum of pure quantum states.  
Further, we can already see intuitively from Fig. \ref{Fig:Dirichlet_Quad} that the concentration parameters offer a powerful way of altering the overall properties of the coefficients.

An ensemble of $D$-dimensional mixed quantum states made from a convex sum of $K$ Haar-random pure states $\ket{\psi_{j}}$ is given by
\begin{equation}
    \rho=\sum_{j=1}^{K}x_{j}\vert\psi_{j}\rangle\langle\psi_{j}\vert,
    \label{eq:D_state_def}
\end{equation}
where the vector $\bx$ is a random variable distributed according to $\text{Dir}(\bx|\alpha)$, having been specialized to the symmetric case $\balpha=\{\alpha,...,\alpha\}$. 
To our knowledge, this ensemble of states was first considered by Mai and Alquier~\cite{mai2017pseudo}, inspired by work on low-rank matrix estimation via machine learning \cite{alquier2015bayesian,cottet20181}.
In light of this, and for convenience in the following analysis, we dub this construction the ``Mai--Alquier'' (MA) distribution.
For our purposes here, we apply the MA label specifically to the case of symmetric Dirichlet weights, but admit any $K\in\mathbb{N}$; thus, for a given dimension $D$, the MA distribution requires specification of two parameters, $\alpha$ and $K$. Since first being defined in \cite{mai2017pseudo}, the MA distribution has been used to analyze data from several quantum optical experiments via Bayesian quantum state tomography~\cite{lukens2020practical, Lu2020b, Lingaraju2021, Alshowkan2021}.

To characterize the properties of an ensemble of states generated according to Eq.~(\ref{eq:D_state_def}), we begin by calculating the mean purity. 
We have chosen purity as our primary metric of interest for several reasons.
The first is that, like the distributions we are analyzing, it is invariant under local rotations, and hence we can abstract away any considerations related to system alignment.  
Further, the purity of a quantum system can be estimated without full state tomography \cite{ekert2002direct}.
Finally, it applies to both individual and composite quantum systems equally well, as opposed to an entanglement metric, which is only applicable to the latter. 
The purity of $\rho$ is given by
\begin{equation}
    \text{Tr}(\rho^{2})=\sum_{j=1}^{K}x_{j}^{2}+2\sum_{j>k}^{K}x_{j}x_{k}F_{j,k},
    \label{eq:purity}
\end{equation}
where $F_{j,k}=\vert\langle\psi_{j}\vert\psi_{k}\rangle\vert^{2}$.
Since $2\sum_{j>k}^{K}x_{j}x_{k}F_{j,k}\ge0$, we see that for a given set of coefficients $\bx$ the minimum purity occurs when all $F_{j,k}=0$, meaning that the states in the sum are orthogonal.
This related case, where all states in Eq.~(\ref{eq:D_state_def}) are orthogonal, was studied by \.{Z}yczkowski in \cite{zyczkowski1999volume} in the context of estimating the volume of the separable states in the space of all mixed quantum states.

Random states of $D$ dimension from the ensemble described in \cite{zyczkowski1999volume} can be generated by taking $K=D$ in Eq.~\eqref{eq:dirichlet_def}, using $\bx$ as the entries in a diagonal matrix, and rotating this matrix by a Haar-random unitary from $U(D)$.  
Many studies have generated random density matrices in this fashion, but of particular interest to applications discussed in Sec.~\ref{sec:nnQST} is \cite{lu2018separability}, which uses this method to generate training data for a machine-learning-based separability-entanglement classifier.
In the Appendix, we characterize the mean purity of the distribution described by \.{Z}yczkowski as a function of the Dirichlet concentration coefficients and compare these results with the Bures, HS, and MA distributions.  
We find that the \.{Z}yczkowski distribution is, as the MA distribution, capable of biasing based on mean purity.  
Also in the Appendix, we compare how well the  \.{Z}yczkowski and MA distributions fit the experimental scenarios of Sec.~\ref{sec:exp}, and find in all three scenarios the MA distribution fits as well or better than the \.{Z}yczkowski distribution, as measured by the Bhattacharyya coefficient \cite{fuchs1999cryptographic}. 
As a final point of comparison, in Sec.~\ref{sec:comparison} we will find that the MA distribution appears to simplify to the HS distribution given the correct choice of concentration parameter, which, to our knowledge, is not possible with the \.{Z}yczkowski distribution. For these reasons, we focus our analysis on the MA distribution in our comparisons with Bures and HS here.

The expectation value of the purity can be found from
\begin{equation}
\begin{aligned}
    \text{E}_{MA}\left[\text{Tr}(\rho^{2})\right]=&\sum_{j=1}^{K}\text{E}\left[x_{j}^{2}\right]+2\sum_{j>k}^{K}\text{E}\left[x_{j}x_{k}F_{j,k}\right]\\
    =& \sum_{j=1}^{K}\text{E}\left[x_{j}^{2}\right]+2\sum_{j>k}^{K}\text{E}\left[x_{j}x_{k}\right]\text{E}\left[F_{j,k}\right]\\
    \end{aligned}
\end{equation}
where we have used the linearity of the expectation operator and 
the fact that the states $\ket{\psi_j}$ are statistically independent of the weight coefficients $x_j$. (When needed for clarity in this and what follows, we apply a subscript $s\in\{MA,B,HS\}$ to expectations and variances in order to distinguish the density matrix distribution over which the quantity is computed.)
For two Haar-randomly chosen pure states, the fidelity is distributed according to \cite{zyczkowski2005average}
\begin{equation}
    P(F)=(D-1)(1-F)^{D-2}.
    \label{eq:fidelity_dist}
\end{equation}
From this we can calculate the first moment as
\begin{equation}
        \text{E}\left[F\right]=\int_{0}^{1}F P(F) \text{d}F = \frac{1}{D},
        \label{eq:mean_fidelity}
\end{equation}
which corresponds with the expression in \cite{zyczkowski2005average}.
The remaining expectation values can be found from Eq.~(\ref{eq:stat}) and the standard relationships
\begin{equation}
\begin{aligned}
    \text{Var}\left[X\right]&=\text{E}\left[X^{2}\right]-E\left[X\right]^{2}\\ \text{Cov}\left[X_{j},X_{k}\right]&=\text{E}\left[X_{j}X_{k}\right]-\text{E}\left[X_{j}\right]\text{E}\left[X_{k}\right].\\
\end{aligned}
\end{equation}
The resulting expectation value of the purity is then given by
\begin{equation}
        \text{E}_{MA}\left[\text{Tr}(\rho^{2})\right]=\frac{D+\alpha(D+K-1)}{D(1+\alpha K)},
    \label{eq:purity_av1}
\end{equation}
where we have evaluated $\sum_{j>k}^{K}=\frac{K^{2}-K}{2}$.

\begin{figure}[!tb]
\centering 
	\includegraphics[width=3.4in]{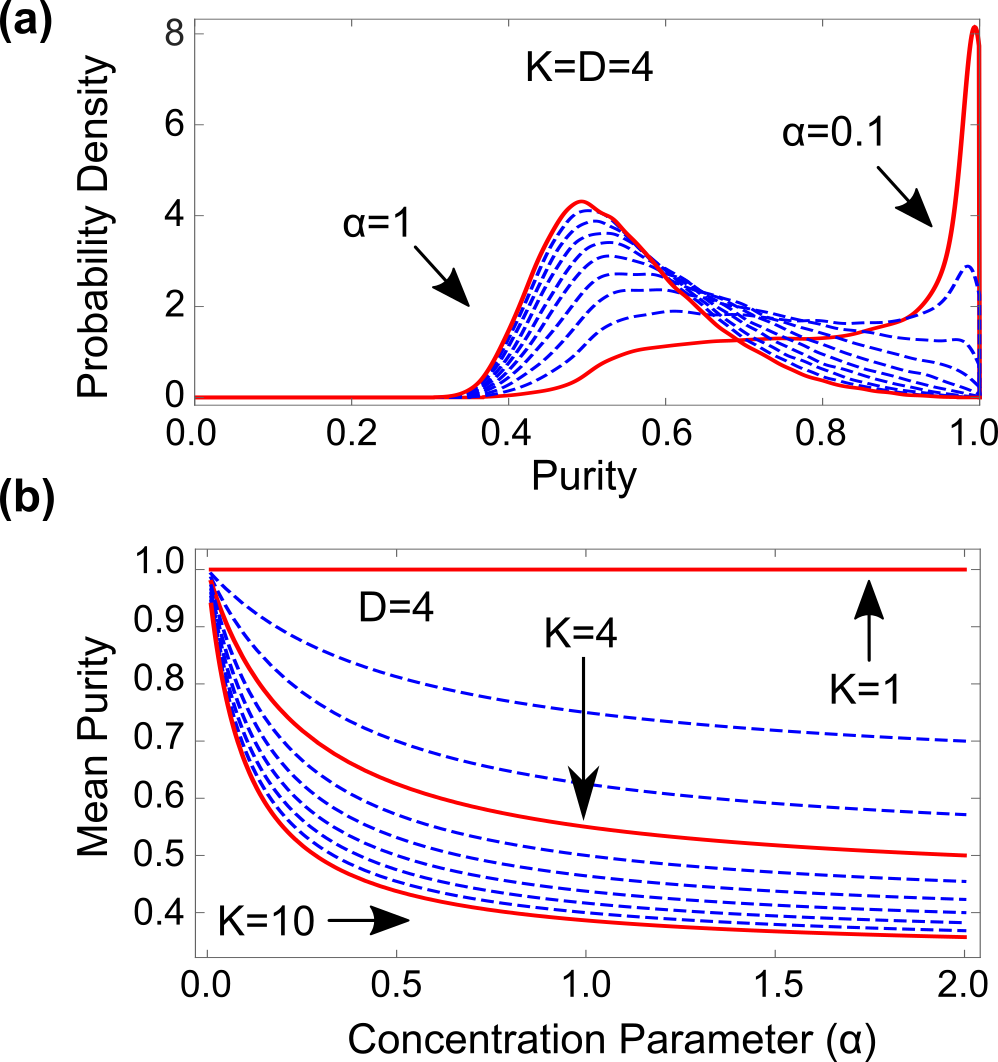}
	\caption{Behavior of the probability density function and mean purity of the MA distribution.  (a) Numerically sampled and smoothed probability density function of MA distributed random states for $K=D=4$.  Ten different curves are shown corresponding to $\alpha=0.1$ to $\alpha=1$ in steps of $0.1$.  The solid (red) lines are $\alpha=0.1$ and $\alpha=1$, with the other values shown as dashed (blue) lines.  (b) Mean purity of MA distributed states for $D=4$ and $K=1$ to $K=10$ in steps of 1.  The solid (red) lines indicate $K=1,4,10$, with the other values shown with dashed (blue) lines.  The $K=4$ curve corresponds to the curves in part (a).}
	\label{Fig:Dirichlet_2}
\end{figure}

It is instructive to consider the behavior of Eq.~(\ref{eq:purity_av1}) in several special cases.
In the limit $\alpha\rightarrow\infty$, the variance of the symmetric Dirichlet distribution vanishes, and all coefficients approach $x_{j}=1/K$.
This limit corresponds to the case explored in both \cite{bengtsson2017geometry,alonso2012ehrenfest}, and we find that for large $\alpha$ our Eq.~(\ref{eq:purity_av1}) reduces to the expression found in \cite{alonso2012ehrenfest}: $\text{E}\left[\text{Tr}(\rho^{2})\right]=(D+K-1)/DK$.
(We will return to the results of \cite{bengtsson2017geometry} in Sec. \ref{sec:comparison} in the context of comparisons with the HS ensemble.) In the opposite limit where all of the $\alpha\rightarrow 0$, the Dirichlet distribution favors a single entry of value $1$ with all others $0$~\cite{Telgarsky2013}, and thus we find a mean purity of unity as all states are rank-1.

The ability to alter the properties of this ensemble of states is further exemplified in Fig. \ref{Fig:Dirichlet_2}(a), where we plot the smoothed probability density function obtained from numerical samples 
of the purity for different $\alpha$ when $K=D=4$ (corresponding to two qubits).
In particular, we show ten values of $\alpha$ in steps of $0.1$ from $\alpha=0.1$ to $\alpha=1$, with these extremal cases plotted in solid red and the intermediate cases in dashed blue. 
Note also that the $\alpha$ values of the solid red lines correspond to those visualized for the $K=3$ distribution in Fig.~\ref{Fig:Dirichlet_Quad}(a,c).

In Fig.~\ref{Fig:Dirichlet_2}(b) we consider the impact of increasing $K$ for a fixed $D$, in other words, making each mixed state in the ensemble a sum of more pure states.
Specifically plotted 
is Eq.~(\ref{eq:purity_av1}) with $D=4$ and $K$ ranging from $1$ to $10$ in single increments, as a function of $\alpha$, with the end cases in solid red.
Further, for ease of comparison with Fig.~\ref{Fig:Dirichlet_2}(a), we also make the $K=D=4$ case in solid red, where the mean of each distribution corresponds to the $y$ value along the $K=4$ line.
As expected, the more terms in the sum of each $\rho$, the more rapidly the states in the ensemble become mixed as $\alpha$ increases.

Finally, we can also place an upper bound on the variance of the purity of $\rho$ using the Bhatia--Davis inequality as \cite{bhatia2000better}
\begin{equation}
    \text{Var}_{MA}\left[\text{Tr}(\rho^{2})\right]\le\frac{\alpha(1+\alpha)(D-1)^{2}(K-1)}{(D+\alpha D K)^{2}}.
\end{equation}
We see for a fixed $D$ and $K$ that in the limit $\alpha\rightarrow 0$ we find $\text{Var}_{MA}\left[\text{Tr}(\rho^{2})\right]\rightarrow 0$, which is consistent with the intuition that all but one element in $\bx$ approaches zero in this limit, meaning that $\rho$ becomes a pure state.
In the opposite limit of $\alpha\rightarrow\infty$, we find for fixed $D$ and $K$ that
\begin{equation}
    \lim_{\alpha\rightarrow\infty}\text{Var}_{MA}\left[\text{Tr}(\rho^{2})\right]\leq\frac{(D-1)^2 (K-1)}{D^2 K^2}.
\end{equation}
This limit includes but does not equal $0$, which is consistent with the interpretation that when $\alpha\rightarrow\infty$, the state will not always approach the completely mixed state.
This is obvious if we consider $K<D$, as it is impossible for the state to become full rank, and thus no set of coefficients could result in an identity matrix.
However, we note that a completely mixed state can be reached consistently for any $\alpha$ and $D$ if we instead take the limit
\begin{equation}
    \lim_{K\rightarrow\infty}\text{Var}_{MA}\left[\text{Tr}(\rho^{2})\right]=0.
\end{equation}
Physically, this can be interpreted to mean that if $\rho$ is composed of a sufficiently large sum of Haar-random pure states, it will always approach the maximally mixed state, regardless of coefficients.

\subsection{Comparisons with standard methods}
\label{sec:comparison}

We begin by reviewing standard methods for generating random quantum states according to the Bures and HS distributions.
While each can in principle be induced through the partial trace on higher dimensional pure states \cite{zyczkowski2001induced}, it is simpler in practice to derive each from the complex Ginibre ensemble \cite{al2010random}, which consists of $D\times D$ complex matrices with independently chosen entries from a standard normal distribution \cite{ginibre1965statistical}. 
A random quantum state $\rho$ from the Bures ensemble can be generated according to 
\begin{equation}
        \rho=\frac{\left(\mathds{1}+U\right)GG^{\dagger}\left(\mathds{1}+U^{\dagger}\right)}{\text{Tr}\left[\left(\mathds{1}+U\right)GG^{\dagger}\left(\mathds{1}+U^{\dagger}\right)\right]},
    \label{eq:Bures}
\end{equation}
and from the Hilbert--Schmidt ensemble by
\begin{equation}
        \rho=\frac{GG^{\dagger}}{\text{Tr}\left(GG^{\dagger} \right)},
    \label{eq:HS}
\end{equation}
where $G$ is a random matrix from the Ginibre ensemble and $U$ is a Haar-distributed random unitary from $U(D)$~\cite{al2010random}.

The average purity of Bures-distributed states was found in \cite{al2010random} and is given by \begin{equation}
    \text{E}_B\left[\text{Tr}(\rho^{2})\right]=\frac{5D^{2}+1}{2D(D^{2}+2)}.
    \label{eq:buresav}
\end{equation}
Similarly, for the HS distribution, the average purity was found to be \cite{zyczkowski2001induced}
\begin{equation}
    \text{E}_{HS}\left[\text{Tr}(\rho^{2})\right]=\frac{2D}{D^{2}+1}.
    \label{eq:HSav}
\end{equation}
We immediately see that \cite{sommers2004statistical}
\begin{equation}
    \text{E}_B\left[\text{Tr}(\rho^{2})\right]>\text{E}_{HS}\left[\text{Tr}(\rho^{2})\right]
\end{equation}
for any dimension $D\ge2$.

\begin{figure}[!tb]
\centering 
	\includegraphics[width=3.4in]{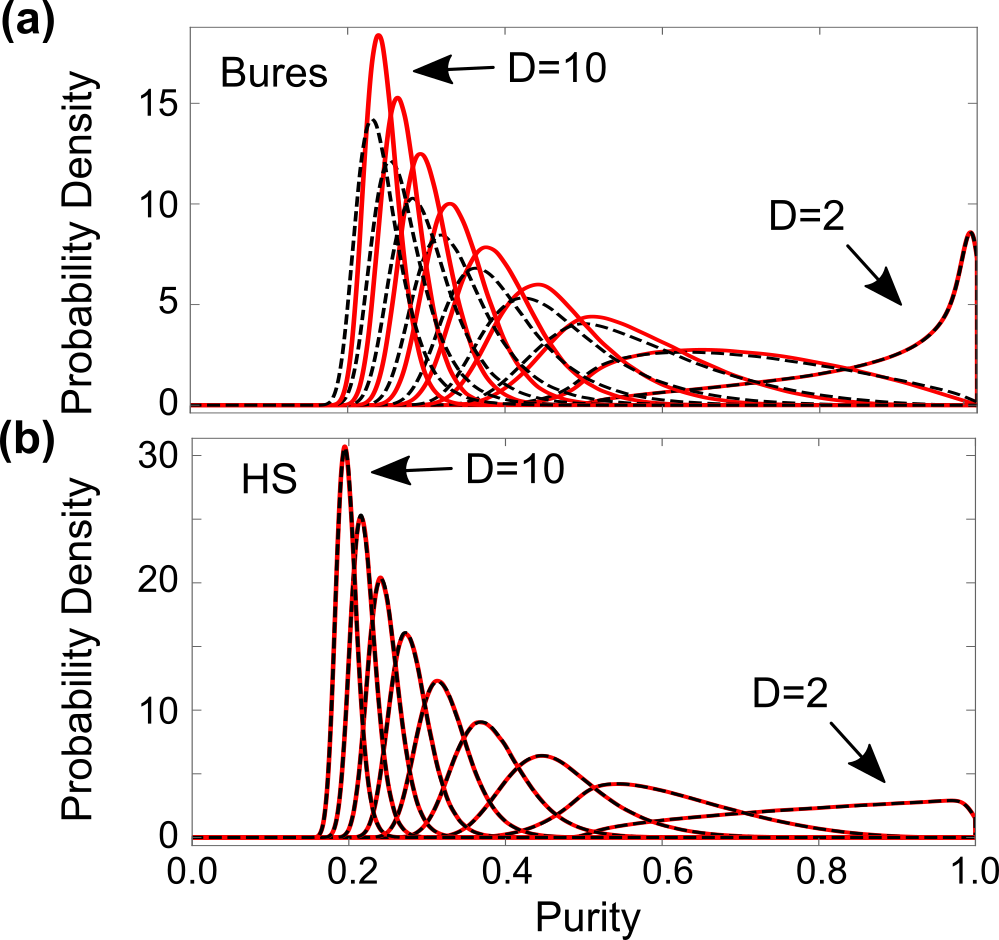}
	\caption{Comparison of the numerically sampled probability distribution of purity for MA, Bures, and HS distributed random quantum states.  Both plots show $D=K=\{2,...,10\}$, and each curve was created using $10^6$ samples.  (a) The Bures distribution is plotted with solid (red) lines, and the corresponding MA distribution with $\alpha=\alpha_{B}$ is plotted with dashed (black) lines.  (b) The HS distribution is plotted with  solid (red) line, with the corresponding MA distribution with $\alpha=\alpha_{HS}$ plotted with dashed (black) lines.  We see complete overlap between the HS and MA distributions when $\alpha=\alpha_{HS}$.}
	\label{Fig:comparison}
\end{figure}

For ease of comparison, we fix $K=D$ in Eq.~(\ref{eq:purity_av1}), which becomes
\begin{equation}
        \text{E}_{MA}\left[\text{Tr}(\rho^{2})\right]_{K=D}=\frac{D+\alpha(2D-1)}{D+\alpha D^{2}}.
    \label{eq:purity_N=D}
\end{equation}
We note that $K=D$ is a natural choice as it is the minimum $K$ capable of producing a full rank state.
From these expressions, we find that $\text{E}_{MA}\left[\text{Tr}(\rho^{2})\right]_{K=D}$ can be made equal to either the Bures or HS averages by setting $\alpha$ equal to the following 
\begin{equation}
    \begin{aligned}
    \alpha_{B}&=\frac{2D-1}{4+D}\\
    \alpha_{HS}&=D.\\
    \end{aligned}
    \label{eq:alpha_reproduce}
\end{equation}
The probability density for the purity of the Dirichlet-distributed mixed states and the Bures and Hilbert--Schmidt ensembles are plotted in Fig.~\ref{Fig:comparison}.
The Bures and Hilbert--Schmidt probability densities are in solid (red) with the $\alpha=\alpha_{B}$ and $\alpha=\alpha_{HS}$ MA ensembles plotted with dashed (black) lines.
The probability densities are shown for $K=D$ and $D\in\{2,...,10\}$: the $D\in\{2,10\}$ cases are labelled, with the others in sequential order.
For all dimensions considered, the tailored MA distribution appears to reproduce the HS ensemble exactly.

For all $D\ge 2$, $\alpha_{HS}>\alpha_{B}$, and hence we arrive at the following inequalities:
\begin{equation}
    \mu_{MA}^{\alpha<\alpha_{B}}>\mu_B>\mu_{HS}>\mu_{MA}^{\alpha>\alpha_{HS}}
    \label{eq:ineq}
\end{equation}
where we have used the shorthand $\mu=\text{E}\left[\text{Tr}\left(\rho^{2}\right)\right]$ and the MA distributions assume $K=D$.
The findings summarized in Eqs.~$(\ref{eq:alpha_reproduce},\ref{eq:ineq})$ represent major contributions of our present investigation, revealing quantitatively how the parameter $\alpha$ of the MA distribution can be tuned to obtain an equal, higher, or lower mean purity compared to well-known fiducial density matrix measures. As we will see below, for many applications the system under investigation may produce states that are nearly pure on average, and hence randomly sampling states with $\alpha<\alpha_{B}$ can bias the distribution to better reflect the states of interest. Such tunability in rank is unavailable from Bures (by definition); for HS, changing to draws of complex-normal $D\times K$ matrices $G$  (with $K<D$) can be used to reduce the rank---and increase purity---of $\rho$ formed using Eq.~(\ref{eq:HS})~\cite{Granade2016}, but this comes at the cost of eliminating full-rank states from consideration altogether. In contrast, adjusting $\alpha$ in the MA ensemble is able to favor pure or mixed states while still providing support over the full $D$-dimensional Hilbert space. 

Finally, as a side note, the $\alpha$ values found to match the mean purities of Bures and HS distributions in Eq.~(\ref{eq:alpha_reproduce}) are almost as interesting for what they are \emph{not} as for what they are. For example, neither case corresponds to $\alpha=1$, despite the fact that $\alpha=1$ gives a uniform distribution of coefficients over the simplex [Fig.~\ref{Fig:Dirichlet_Quad}(a)] and therefore ostensibly may seem the default choice for a uniform density matrix distribution. 
At the other extreme, nor does the  limit $\alpha\rightarrow\infty$ give mean purities commensurate with Bures or HS. Indeed this latter limit corresponds to the sum of $D$ evenly weighted Haar-random pure states, which was noted in \cite{bengtsson2017geometry} to differ from the HS ensemble in general; thus our conclusion $\alpha_{HS}\neq\infty$ is consistent with these findings as well.

\section{Applications}
\begin{figure}[!tb]
\centering 
	\includegraphics[width=.8\columnwidth]{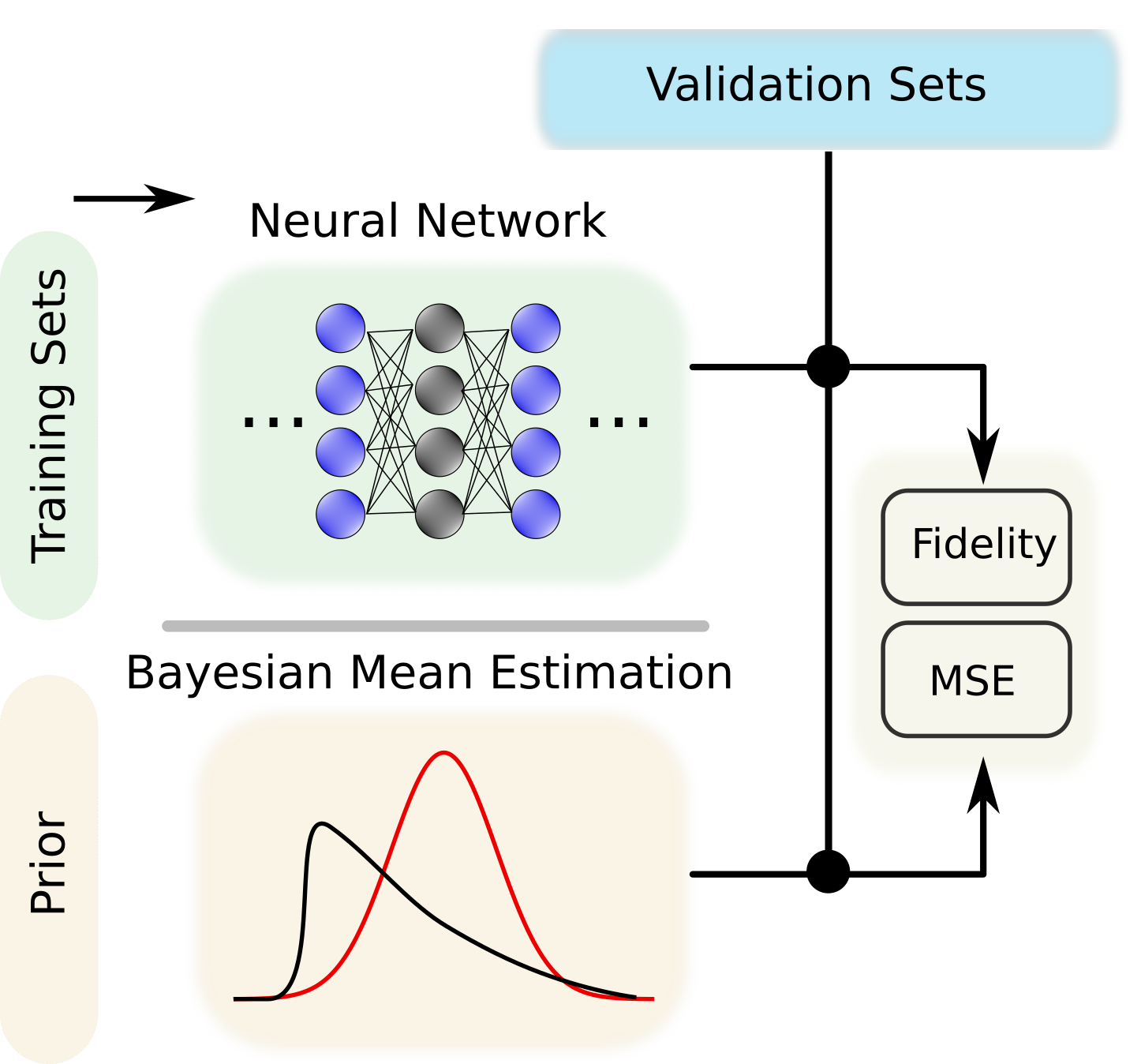}
	\caption{Two advanced methods for reconstructing quantum states: neural network (top) and Bayesian mean estimation (bottom). Both methods require a defined distribution as input: the training set for the neural network, and prior for BME.  In both methods, fidelity and mean-squared error (MSE) between the target and reconstructed quantum states are evaluated for common validation sets.}
	\label{Fig:application-intro}
\end{figure}
We now explore 
applications that show measurable performance improvements when substituting the MA distribution for either the Bures or HS distributions.  In particular, we consider two different methods for reconstructing quantum states from measured tomography data as shown in Fig. \ref{Fig:application-intro}.  The first method reconstructs states using pre-trained neural networks.  In Sec.~\ref{sec:nnQST}, we study the impact of the chosen training set on the fidelity of the reconstruction.  The second approach we consider for state reconstruction is Bayesian mean estimation (BME).  In Sec.~\ref{sec:Bayesian}, we examine the significance of using a carefully selected prior distribution for the rate of convergence of a BME-based reconstruction method. 

So that we can study the efficacy of these techniques in a range of situations, we have opted to perform state reconstruction as a function of the number of measurements performed.  In particular, we perform reconstruction of ($D=4$)-dimensional states using simulated datasets from random Pauli measurements (one measurement per each random basis selection) with the total number of measurements ranging from 1 to 1000.
This approach allows us to view the disparity in performance both in the high-statistical noise regime where very little data is available and in the more ideal asymptotic case where the observed frequencies approach the actual probability distributions. 
The low-count regime is of particular interest for tomography of high-dimensional systems where the size and complexity of the system may limit the amount of measurements that can be feasibly performed. 

In each section, we perform numerical simulations using  
sets of randomly sampled quantum states.  For training, we utilize full 1000-measurement datasets from $10^5$ randomly chosen states from a given metric: Fubini--Study (pure states), Bures, HS, or MA of a particular $\alpha$.   Pure states of dimension $D=4$ were sampled by taking the first column of a Haar-distributed unitary from $U(D)$.  We sample the Bures and HS distributions using 
Eqs. \eqref{eq:Bures} and \eqref{eq:HS}. Validation is then performed using 2000 randomly chosen states from the metric under consideration (which need not match the training set), where we consider total measurement numbers of $M\in\{1, 15, 25, 50, 75, 100, 200, 400, 600, 800, 1000\}$. The Bayesian approach does not utilize a training set (the analogue functionality is realized by the mathematically specified prior distribution), but performance is tested using the same validation sets as the neural network cases.

\subsection{Neural network quantum state tomography}
\label{sec:nnQST}
\begin{figure}[!b]
\centering 
	\includegraphics[width=.9\columnwidth]{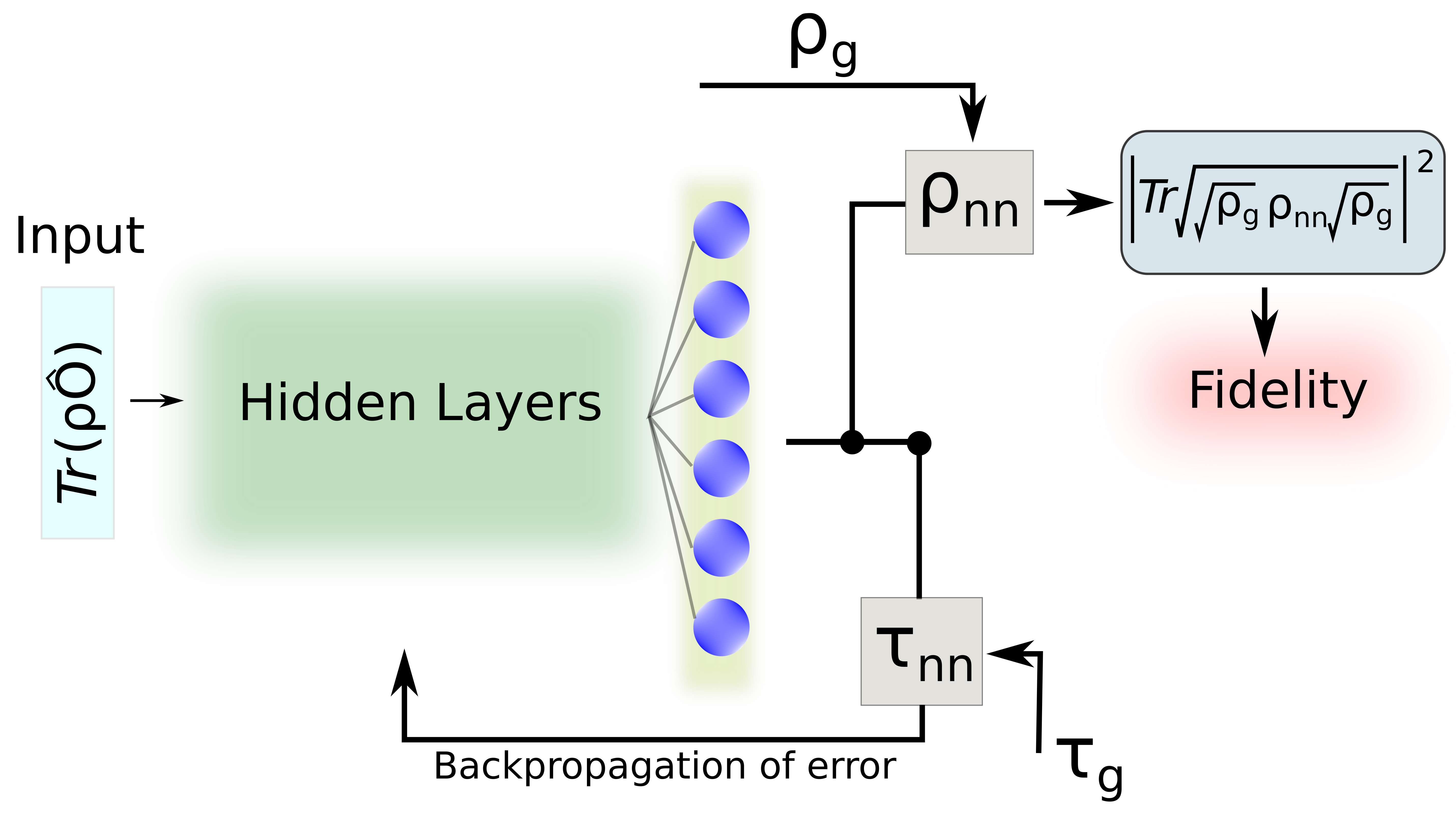}
	\caption{Architecture of the neural network. $\rho_g$ and $\rho_{nn}$ represent the ground truth and predicted density matrices. Similarly, $\tau_g$, and $\tau_{nn}$, respectively are the ground truth and predicted tau-vectors.}
	\label{Fig:Architecure_NN}
\end{figure}
Machine learning has found broad applicability in quantum information science in topic areas as diverse as experiment design \cite{melnikov2018active}, state classification \cite{sentis2015quantum,harney2020entanglement,lu2018separability}, and even studies on quantum foundations \cite{bharti2020machine}.  Recently, several studies have applied machine-learning methods to quantum state reconstruction \cite{lu2018separability,lohani2020machine,danaci2021machine,lohani2020experimental,ahmed2020classification}.  Here we consider how well a machine-learning-based quantum state tomography system can reconstruct states of one distribution when trained on another.  Specifically, we will first consider  
reconstructing pure two-qubit quantum states by training a network exclusively on either (i) Haar-random pure states, (ii) randomly generated states from the Bures distribution, or (iii) randomly generated states from the HS distribution.  Further, we will demonstrate the versatility of the MA distribution by separately training a network on randomly generated states from ten different MA distributions ($\alpha$ from 0.1 to 1 in steps of 0.1), and we will show how well each of these networks reconstructs pure-, HS-, and Bures-distributed validation states.  We note that a previous study has considered the issue of matching training distributions to validation distributions and dealt with it using a pipeline of neural networks trained on different distributions \cite{danaci2021machine}.  Instead, in this study we consider the generalizability of a single neural network for reconstructing states from the entire Hilbert space.

The architecture of the neural network we consider is shown in Fig. \ref{Fig:Architecure_NN}.
The network consists of an input layer that receives the measured tomography values, $\Tr(\rho \hat{O})$, which connects to a convolutional layer with a kernel size of (2, 2), stride lengths of 1, ReLU as an activation function, and filters of size 64. Then we attach a max-pooling unit with a pool-size of (2, 2), stride of length 2, and the ``valid" padding, which is, again, followed by another convolutional layer with the same configuration as mentioned before. Next, we implement a flattening layer, which is then fully connected to a dense layer with 3000 neurons using the ReLU activation function.   We then apply a dropout unit with a $50\%$ dropout rate, followed by another dense layer with 1200 neurons using a ReLU activation function, followed by a dropout unit with the same rate. After this, we fully connect another dense layer with a linear activation as shown by blue circles in Fig. \ref{Fig:Architecure_NN}. The prediction made at this layer (blue circles in Fig. \ref{Fig:Architecure_NN}) is branched into two pipelines. The first pipeline evaluates the $\tau_{nn}$-vectors and compares them with the ground truth values, $\tau_{g}$. Note that the target $\tau_g$-vectors are obtained from the Cholesky decomposition of the target density matrices ($\rho_{g}$). In the two-qubit scenario, for any random quantum state, $\rho_g$ = $\xi\xi^\dagger$, where 
\begin{equation}
\begin{aligned}
    &\xi\,=\,
        \begin{bmatrix}
        \tau_0 & 0& 0& 0\\
        \tau_4+i\tau_5 & \tau_1 &0 &0\\
        \tau_{10}+i\tau_{11} & \tau_6+i\tau_7 &\tau_2 &0\\
        \tau_{14}+i\tau_{15} & \tau_{12}+i\tau_{13} &\tau_8+i\tau_9 &\tau_3\\
        \end{bmatrix}, \\
\end{aligned}
\label{eqn:cholesky_T}
\end{equation}
which can be rearranged as the vector
\begin{equation}
\begin{aligned}
\quad \xi \rightarrow \tau = (\tau_0, \, \tau_1,\, \tau_2,\, \tau_3,\,  .. .. ..,\, \tau_{15}).
\end{aligned}
\label{eqn:cholesky_T_to_t}
\end{equation}
Consequently, the mean-squared loss between the predicted and the target $\tau$-vectors is obtained and back-propagated for the learning, while the second pipeline makes predictions for the corresponding density matrices from the predicted $\tau_{nn}$-vectors. At the end, the network compares the predicted density matrices ($\rho_{nn}$) with the ground truth states, and evaluates the fidelity
\begin{equation}
\label{eq:fid}
 F(\rho_{nn},\rho_g) = \left[ \Tr\sqrt{\sqrt{\rho_{g}}\rho_{nn}\sqrt{\rho_{g}}}\right]^2.   
\end{equation}
Ultimately, the final network settings are chosen to maximize the average 
fidelity of the reconstructed states to the ground truth.

\begin{figure}[!tb]
\centering 
	\includegraphics[width=3.4in]{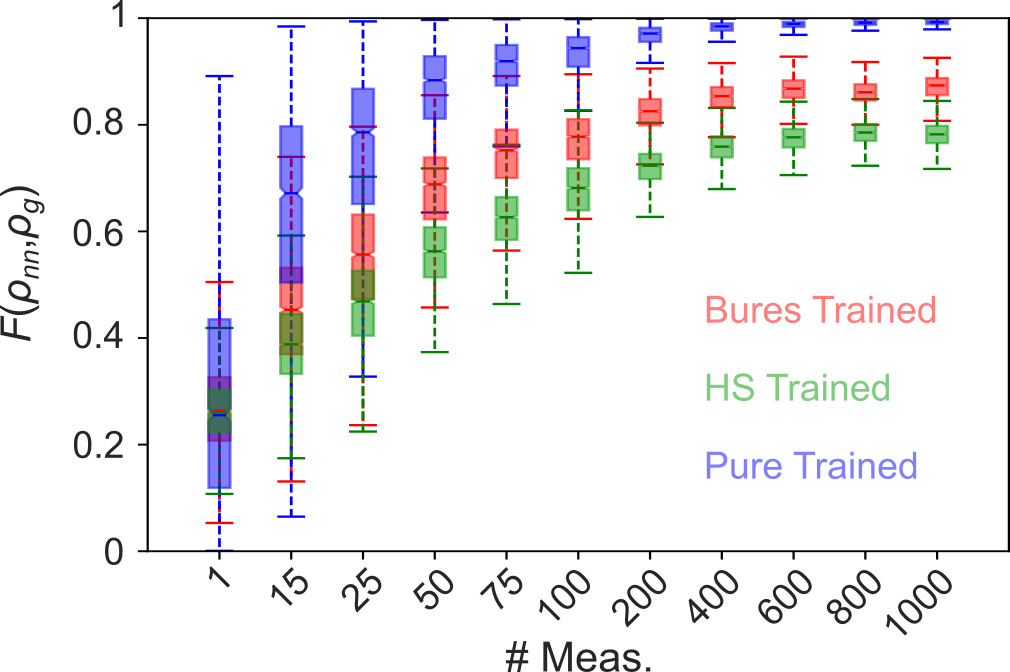}
	\caption{Reconstructing pure states from networks separately trained with either Haar-random pure states, Bures-drawn mixed states, or HS-drawn mixed states.}
	\label{Fig:Reconstruct_Pure}
\end{figure}

To illustrate the proof of concept, we reconstruct two-qubit quantum states using networks separately trained using each of the ensembles introduced above.
Each network is trained for 75 epochs at a learning rate of 0.008. 
Once the networks are trained, we make predictions for a validation set of pure quantum states that are unknown to the networks. 
The reconstruction fidelities for the predicted quantum states for various measurement scenarios are shown in Fig. \ref{Fig:Reconstruct_Pure}. 
We find that the network trained with pure states quickly reconstructs the validation set with near unit fidelity as shown by blue box plots.
However, the networks separately trained with Bures (red box) and HS (green box) states struggle to reconstruct the validation set with reconstruction fidelities leveling off around $0.87$, and $0.79$, respectively. 
For this and all subsequent box plots below, the notch gives the median and each box encloses $[Q_1, Q_3]$, while the whiskers extend from $Q_1 - 1.5(Q_3-Q_1)$ to $Q_3 + 1.5(Q_3-Q_1)$, where $Q_1$ and $Q_3$ are the first and third quartiles. 
At first, this is somewhat surprising, as the pure states are a subset of the mixed states. Hence, a naive assumption is that a system capable of reconstructing mixed states with high fidelity would also reconstruct pure states with similar fidelity.
Therefore, we are faced with a problem of network generalization: the reconstruction fidelity of a neural network depends heavily on the distribution used to train it.
One possible way to overcome this generalization issue is to train multiple neural networks for different situations, as we performed in \cite{lohani2020machine}, or even to combine them together into a pipeline, as was done in \cite{danaci2021machine}. 
Instead, we attempt to solve the same problem by biasing the training distribution to better cover the Hilbert space through proper tuning of the MA distribution. 

\begin{figure}[!tb]
\centering 
	\includegraphics[width=3.4in]{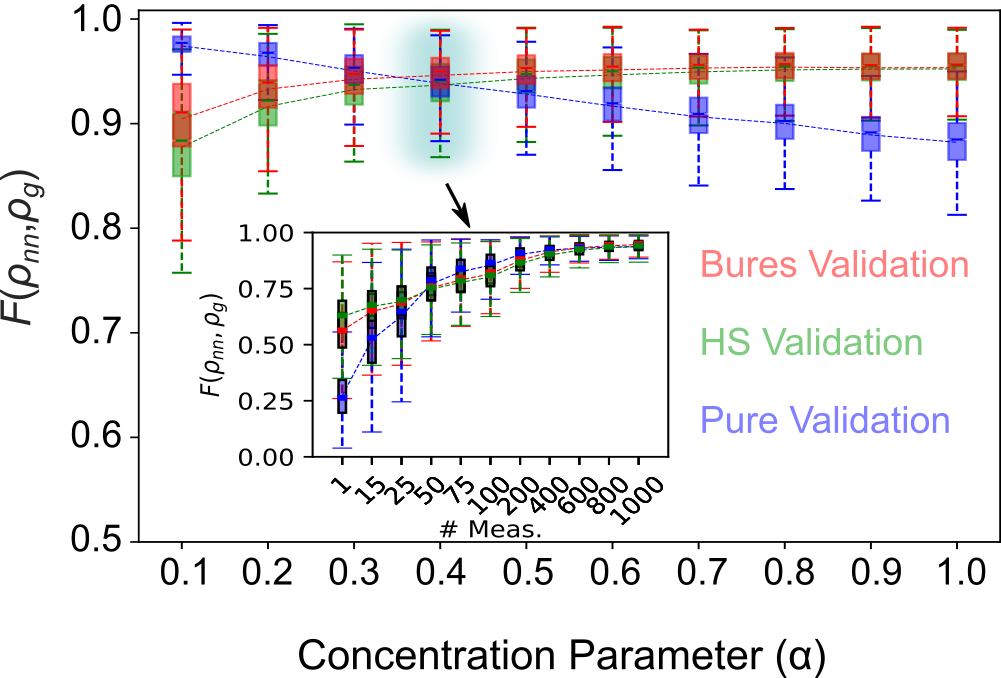}
	\caption{Reconstruction fidelity versus various concentration parameters ($\alpha$). The fidelity at various measurements with the network trained at $\alpha\,=\,0.4$ is shown in the inset, which corresponds to Fig. \ref{Fig:Reconstruct_Pure}.}
	\label{Fig:NN_Alpha}
\end{figure}
In Fig. \ref{Fig:NN_Alpha}, we vary $\alpha$  of the MA distribution from $0.1$ to $1.0$ in steps of $0.1$ and sample training sets for 1000 random Pauli-measurements. 
For each $\alpha$, we separately train a network on states randomly sampled from the corresponding MA distribution. Once the network is trained, we reconstruct the validation sets containing Bures-, HS-, and Haar-distributed pure states individually. 
The reconstruction fidelities for Bures and HS states gradually increase with an increase in $\alpha$ (red and green box plots), whereas the reconstruction fidelities for Haar-random pure states slowly decrease with an increase in $\alpha$ (blue box plots). 
However, reconstruction fidelities for the three cases approximately coalesce when training states are sampled from a distribution with $\alpha \approx 0.4$.
The difference in reconstruction fidelity seen in Fig. \ref{Fig:NN_Alpha} at $\alpha=0.1$ and $\alpha=1$ for different validation sets stresses the importance of carefully selecting training sets for machine-learning-based reconstruction techniques.  
For comparison, we also include an inset in Fig. \ref{Fig:NN_Alpha} in the same form as Fig. \ref{Fig:Reconstruct_Pure}. 
We see in the inset that reconstruction of all three validation sets converge, unlike the situation in Fig. \ref{Fig:Reconstruct_Pure}.

This investigation highlights that by tuning the degree of sparsity of the Dirichlet distribution, we are able to hedge our bets on accurate state estimation. 
A neural network trained with a parameter setting of $\alpha=0.4$ attains significantly higher reconstruction fidelities for validation on pure states compared to a neural network trained on Bures.
Further, as we see in Fig. \ref{Fig:NN_Alpha}, this gain in pure state reconstruction fidelity (blue boxes), comes with a relatively minor reduction in Bures and HS reconstruction fidelity (red and green boxes, respectively).
Hence, through proper tuning of the training set, we are able to create a neural network reconstruction system with significantly more generality than one trained on Bures, HS, or pure states alone. 

\subsection{Bayesian quantum state tomography}
\label{sec:Bayesian}
Here, we continue to explore inference for the same datasets, but now employ Bayesian mean estimation (BME)~\cite{MacKay2003, Blume2010}, ultimately finding similar advantages for the custom states as in the neural network cases. In contrast to alternative approaches in quantum state tomography, Bayesian inference defines a complete probability distribution for $\rho$, utilizing Bayes' theorem to combine prior knowledge and experimental results within a single consistent framework. Bayesian tomography enjoys several appealing features, including automatic uncertainty quantification and the return of reliable estimates under any measurement conditions~\cite{Blume2010}.
Moreover, the mean of the posterior distribution is optimal in that it minimizes the mean-squared error with respect to the ground truth for any number of measurements~\cite{Robert1999}---hence the emphasis on ``mean'' in BME. Although BME remains fairly uncommon in quantum state tomography, due in large part to the computational difficulties associated with high-dimensional integrals, several practical Monte Carlo approaches for quantum BME~\cite{Seah2015,Granade2016,Williams2017,mai2017pseudo,lukens2020practical} have appeared following the initial proposal~\cite{Blume2010}, making it an increasingly more attractive prospect in quantum state estimation.

Formally, in the Bayesian viewpoint we assign a probability distribution to $\bw$, a vector of all parameters necessary to define a given density matrix $\rho(\bw)$. The length and content of $\bw$ can vary depending on the chosen parameterization, but it is assumed that all allowed values produce a physical $\rho(\bw)$, i.e., unit trace, Hermitian, and positive semidefinite. According to Bayes’ rule, the posterior probability density for $\bw$ following an experiment can be written as
\begin{equation}
\label{j1}
\pi(\bw) = \frac{1}{\mathcal{Z}} L_\bbD(\bw) \pi_0(\bw)
\end{equation}
where the likelihood $L_\bbD(\bw)\propto P(\bbD|\bw)$ (the probability of observing dataset $\bbD$ given parameters $\bw$), $\pi_0(\bw)$ is the prior on $\bw$, and $\mathcal{Z}$ is a normalizing constant (which does not need to be determined in the following examples). Considering a collection of $M$ measurements on a repeatedly prepared input state, with each result described by an operator $\Lambda_m$, the likelihood can be written as $L_\bbD(\bw) = \prod_{m=1}^M \Tr \rho(\bw) \Lambda_m$; in the special case of projective measurements (such as the Pauli measurements in our example), this becomes
\begin{equation}
\label{j2}
L_\bbD(\bw) = \prod_{m=1}^M \braket{\psi_m|\rho(\bw)|\psi_m},
\end{equation}
with $\ket{\psi_m}$ the eigenstate observed in measurement $m$.

Of primary interest for our present purposes, however, is the prior $\pi_0(\bw)$. Just as the training set in the neural network case specifies the realm of states expected in subsequent experiments, $\pi_0(\bw)$ expresses all \emph{a priori} assumptions about the system under test in the form of an explicit probability distribution over states $\rho(\bw)$. As frequently emphasized in BME, the exact functional form of any well-chosen prior---e.g., one which assigns reasonable probability to all states in the Hilbert space---will have minimal impact on the posterior $\pi(\bw)$ in the limit $M\rightarrow\infty$, since the likelihood $L_\bbD(\bw)$ then dominates. However, $\pi_0(\bw)$ can have a profound impact in the case of few measurements, such as the conditions explored here.

Given the neural network findings in Fig.~\ref{Fig:NN_Alpha}, we focus our comparison specifically on priors corresponding to either the Bures, HS, or MA ($K=4$, $\alpha=0.4$) distribution.
For each option, we select a parameterization amenable to numerical sampling. 
Bures states can be represented by the vector $\bw=(w_1,…,w_{2D^2})$, with each element independently distributed according to a complex standard normal distribution $w_k \sim \mathcal{CN}(0,1)$. We assign $D^2$ of the components to populate the $D\times D$ Ginibre matrix $G$ in Eq.~(\ref{eq:Bures}); the remaining $D^2$ elements comprise a second, independent Ginibre matrix which is then fed into the algorithm of~\cite{Mezzadri2007} to produce the Haar-random unitary $U$ also required in Eq.~(\ref{eq:Bures}). Thus the Bures prior can be written as $\pi_0(\bw) \propto \prod_{k=1}^{2D^2} e^{-\frac{1}{2}|w_k|^2}$. Since HS states can be represented using a single Ginibre draw [Eq.~\eqref{eq:HS}], we can shorten $\bw$ to a length-$D^2$ vector and define the HS prior as $\pi_0(\bw) \propto \prod_{k=1}^{D^2} e^{-\frac{1}{2}|w_k|^2}$.

For the MA prior, we take $\bw=(\bu,\bv_1,...,\bv_D)$, where $\bu=(u_1,...,u_D)$ is a real $D$-dimensional vector of positive scalars, and $\bv_k$ is a $D$-dimensional complex vector. The prior follows $\pi_0(\bw) \propto \prod_{k=1}^D u_k^{\alpha-1} e^{-u_k} e^{-\frac{1}{2} \bv_k^\dagger \bv_k}$, and we define $\rho(\bw)$ according to
\begin{equation}
\rho(\bw) = \sum_{k=1}^D \left(\frac{u_k}{\sum_l u_l} \right) \frac{\bv_k \bv_k^\dagger}{|\bv_k|^2}.
\label{eq:MAparam}
\end{equation}
This equation is equivalent to Eq.~(\ref{eq:D_state_def}) expressed in the computational basis, though for convenience we now utilize unnormalized parameters; our combined gamma and complex-normal prior ensures that the normalized entities $\bu/\sum_l u_l$ and $\bv_k/|\bv_k|$ are Dirichlet- and Haar-distributed, respectively, as required~\cite{mai2017pseudo}.

For either prior, the Bayesian mean estimate for $\rho$ is
\begin{equation}
\label{j3}
\rho_{BME} = \int d\bw\, \pi(\bw) \rho(\bw),
\end{equation}
which we compare to the ground truth $\rho_g$ by fidelity $F(\rho_{BME},\rho_g)$ [Eq.~\eqref{eq:fid}]. Computation of Eq.~(\ref{j3}) is effected utilizing preconditioned Crank—Nicolson (pCN) Markov chain Monte Carlo (MCMC) techniques~\cite{Cotter2013} recently introduced to quantum state tomography~\cite{lukens2020practical}, with the only modification being an improved, reversible proposal distribution for the gamma random variables $u_k$ based on \cite{Lewis1986}. 
A total of $2^{10}$ samples of $\bw$ are kept in estimating Eq.~(\ref{j3}), and a thinning value of $2^8$ was found sufficient for convergence in fidelity, thus corresponding to total MCMC chain lengths of $2^{18}$.

\begin{figure}[tb!]
\centering 
	\includegraphics[width=3.4in]{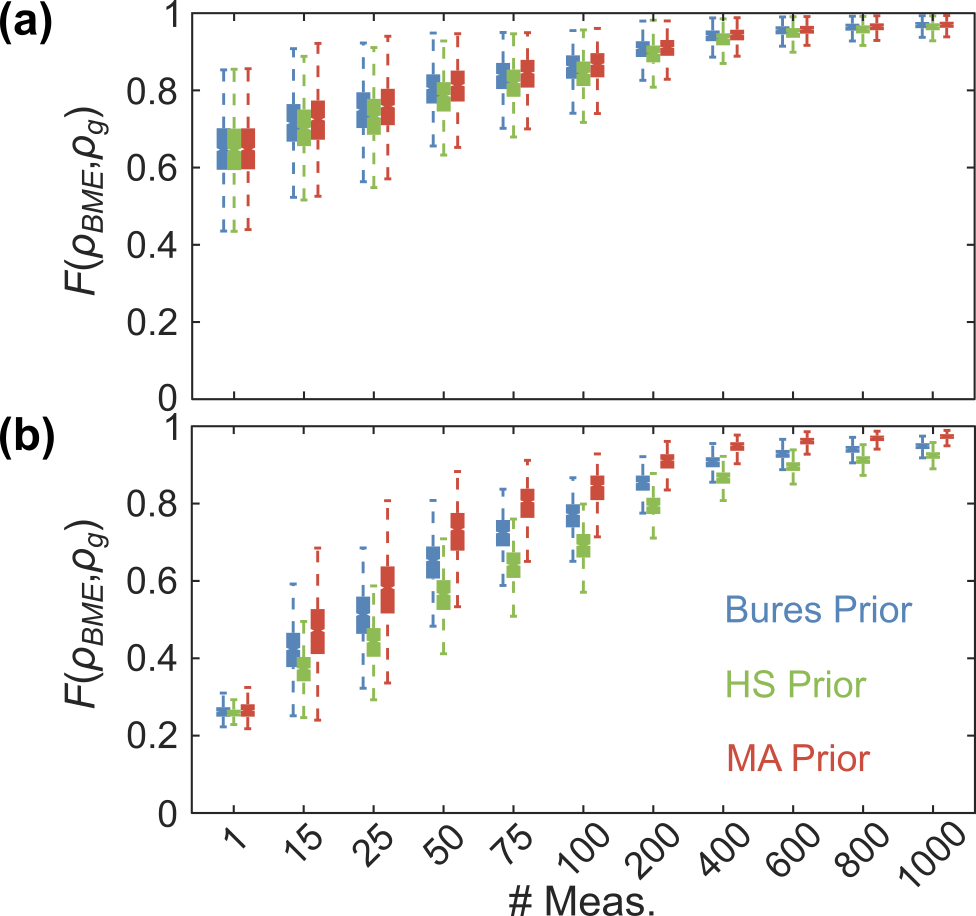}
	\caption{Bayesian inference fidelity for simulated two-qubit experiments. (a) Validation set drawn from Bures metric. (b) Validation set drawn from Fubini--Study metric. Inference is performed using either the Bures, HS, or MA prior (with $K=D$ and $\alpha=0.4$), for a specified number of random Pauli measurements.}
	\label{figJ1}
\end{figure}

The distribution of fidelities $F(\rho_{BME},\rho_g)$ obtained for several sets of quantum states appear in Fig.~\ref{figJ1}, 
plotted as a function of $M$: Fig.~\ref{figJ1}(a) corresponds to ground truth draws (i.e., a validation set) from the Bures distribution, and Fig.~\ref{figJ1}(b) to pure states. 
The fidelities increase rapidly with measurements $M$, and 
all three  priors produce nearly identical distributions for the Bures ground truth states, whereas the MA prior displays a noticeable edge over both Bures and HS when the ground truth draws are pure. 

To further quantify this comparison, we can take advantage of theoretical optimality guarantees offered by the Bayesian mean estimate $\rho_{BME}$, namely, that $\rho_{BME}$ minimizes the expected value of every operational divergence, a type of reward-scheme--motivated metric quantifying the closeness of an estimator to the ground truth $\rho_g$~\cite{Blume2010}. While fidelity does not correspond to such an operational divergence, the squared Frobenius distance $D_F^2 (A,B) = \lVert A-B \rVert_F^2 = \Tr [(A-B)^\dagger (A-B)]$ does. Accordingly, if we define a mean squared error (MSE) according to $ \braket{D_F^2 (\tilde{\rho},\rho_g)}$---the average squared distance over all validation states for a particular type of estimator $\tilde{\rho}$---$\tilde{\rho} = \rho_{BME}$ is guaranteed to minimize MSE, for any number of measurements, provided that the prior $\pi_0(\bx)$ used matches the actual distribution from which $\rho_g$ is drawn.

\begin{figure}[tb!]
\centering 
	\includegraphics[width=3.4in]{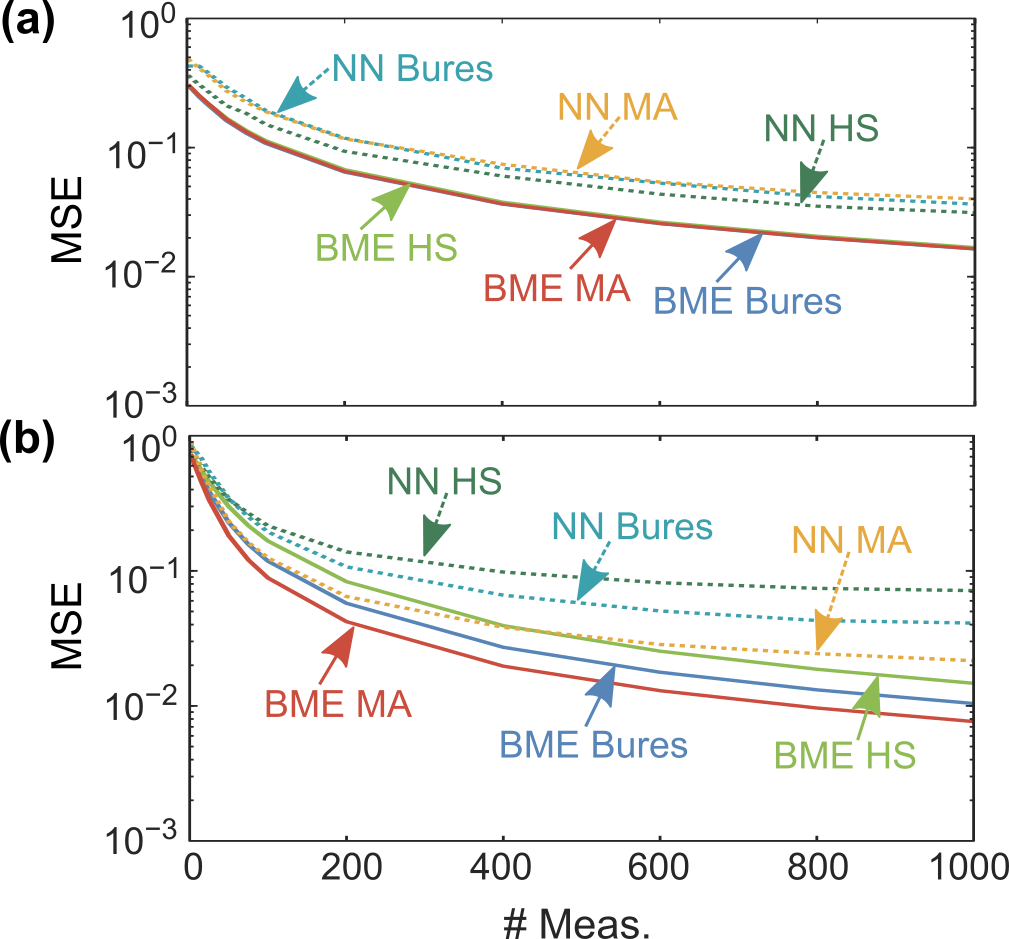}
	\caption{Mean-squared Frobenius error $\braket{D_F^2(\tilde{\rho},\rho_g)}$ for each estimation procedure. Simulated datasets from (a) Bures-drawn states and (b) Haar-random pure states. BME (neural network) results corresponding to either the Bures or MA priors (training sets) are plotted.}
	\label{figJ2}
\end{figure}

Calculating MSE for all examples in Fig.~\ref{figJ1}, as well as for the  outputs of the Bures-, HS-, and MA-trained neural networks of Sec.~\ref{sec:nnQST} with the same input data, we obtain the results in Fig.~\ref{figJ2}. As expected, the BME results based on a Bures prior attain the smallest error of all inference procedures when the validation states are themselves Bures-distributed [Fig.~\ref{figJ2}(a)]. 
Yet the MA prior with $\alpha=0.4$ results prove extremely close and are difficult to distinguish on this scale.
However, when the validation set is restricted to pure states, the MA prior and MA-trained network improve on their Bures and HS counterparts in realizing more accurate estimates [Fig.~\ref{figJ2}(b)]. Importantly, in this latter case, none of the BME priors matches the actual validation distribution, so that MSE optimality does not apply to any curve here, in contrast to the case ``BME Bures'' in (a). As a surprising side note, the neural network trained on HS realizes lower MSE on a Bures validation set than the network actually trained on Bures directly [Fig.~\ref{figJ2}(a)]; the cause of this is unknown, but we suspect it results from the fact that the neural network was trained to maximize fidelity rather than minimize the Frobenius error, for which performance is not guaranteed.

\section{Comparisons with experimental scenarios}
\label{sec:exp}
In the previous section, our use of simulated datasets allowed us to test the performance of the MA distribution directly against fiducial density matrix distributions.
Yet in practice, a quantum system may produce states whose distribution deviates significantly from standard measures such as Fubini--Study, Bures, and HS. Therefore in this section, we will apply the tunability of the MA distribution toward matching the purity distributions of actual experimental systems, specifically 
a cloud-accessed quantum computer (IBM Q) and a commercial polarization-entangled photon source.  

Our primary metric of comparison will be the Bhattacharyya coefficient \cite{fuchs1999cryptographic}.  
We will calculate the Bhattacharyya coefficient from numerically sampled histograms of $100$ bins normalized such that the total sums to unity.
The number of sampled states used to generate these histograms will vary by scenario, and is explicitly mentioned in each section.  
To denote the vector of histogram bin heights, we adopt the notation of $\text{h}_{s}$, where $s\in\{MA,B,HS,IBM1, IBM2,EPS\}$; the first three are for the MA, Bures, and HS distributions, respectively, and the IBM1, IBM2, and EPS subscripts refer to the measured data in Figs. \ref{Fig:ibmq-purity}, \ref{Fig:ibmq-purity2}, and \ref{Fig:EPS-purity}, respectively.
We calculate the Bhattacharyya coefficient from
\begin{equation}
    \mathcal{B}\left(\text{h}_{j},\text{h}_{k}\right)=\sum_{l=1}^{100}\sqrt{h_{j}(l)h_{k}(l)}.
\end{equation}
The Bhattacharyya coefficient between identical histograms is $1$, and two completely nonoverlapping (orthogonal) histograms has a coefficient of $0$.
While the Bhattacharyya coefficient in this discrete form can depend on how many bins we separate the data into, we find that $100$ is sufficient to capture the essential features we want to exemplify.

\subsection{IBM Quantum Computer}
\begin{figure}[!tb]
\centering 
	\includegraphics[width=\columnwidth]{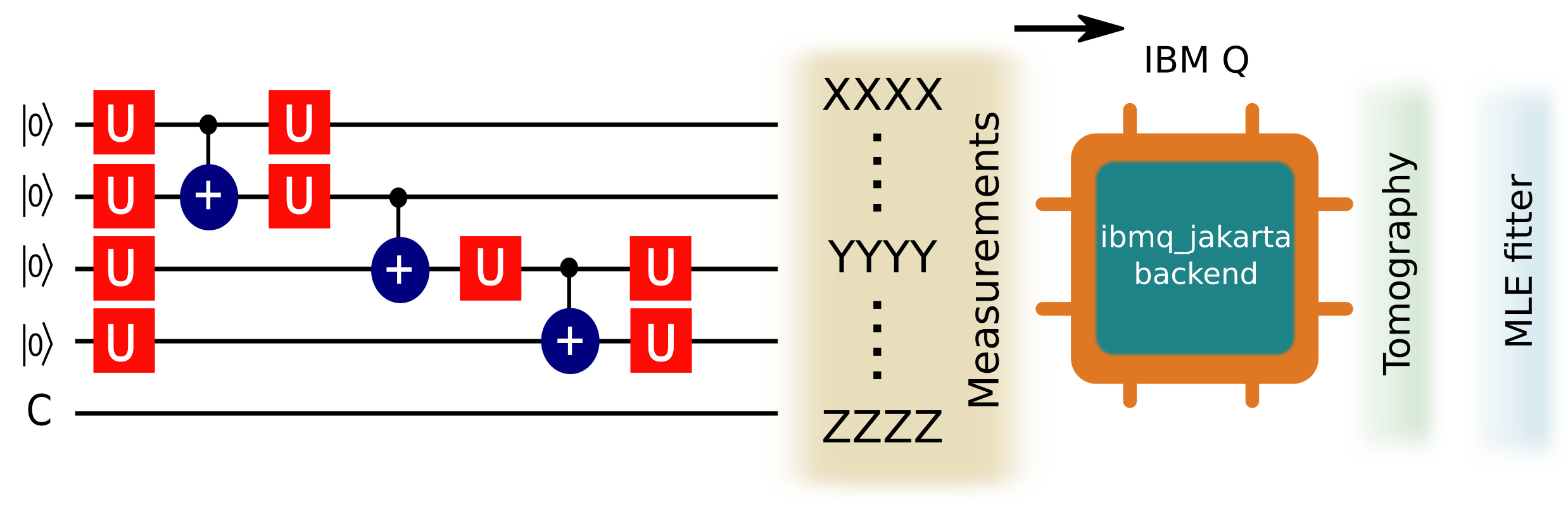}
\caption{IBM Q tomography setup. An arbitrary four-qubit quantum circuit generates random quantum states on a seven-qubit computer ($ibmq\_jakarta$), followed by Pauli measurements. MLE is used to reconstruct the quantum states.}
\label{Fig:ibmq-intro}
\end{figure}
To illustrate the proof of concept, we implement two different scenarios on an IBM  quantum computer that result in distributions of states with significantly different mean purity.
First, we assume a simple approach and initialize a four-qubit quantum state with all qubits in the $|0\rangle$ state and perform tomography without intermediary gates.
The intention of this scenario is to obtain a distribution of states with very high purity in order to mimic the operation of an ``ideal" system.

In the second scenario, we operate on the four initial qubits with a quantum circuit consisting of nine $U$ gates and three $CNOT$ gates, as shown in Fig. \ref{Fig:ibmq-intro}.
The $U$ gates can be represented in matrix form as
\begin{equation}
    U(\theta,\phi, \lambda) = \begin{bmatrix}
                                    \cos(\frac{\theta}{2}) & - e^{i\lambda}\sin(\frac{\theta}{2})\\
                                    e^{i\phi}\sin(\frac{\theta}{2}) &  e^{i(\phi+\lambda)}\cos(\frac{\theta}{2})
                                \end{bmatrix}.
\label{eqn:u-gate-2}
\end{equation}
To generate a distribution of random states, we randomly select the values of $(\theta,\phi, \lambda)$ for the $U$ gates such that $(\theta, \phi, \lambda) \in 2\pi \mathcal{N}(0, 1)$, where $\mathcal{N}(0, 1)$ is the standard normal distribution. 

We perform quantum state tomography using 81 quantum circuits that project on all combinations of the four local Pauli bases $\{X,Y,Z\}_{1}\otimes\{X,Y,Z\}_{2}\otimes\{X,Y,Z\}_{3}\otimes\{X,Y,Z\}_{4}$.
All measurements were executed over the cloud with a seven qubit NISQ-era quantum computer, $ibmq\_jakarta$  \cite{Qiskit}. 
To reduce statistical noise in the IBM Q measurement results, we execute circuits for 5000 shots.  Further, we implement a measurement correction fitter for a full calibration with the method ``least\_squares,'' and perform maximum likelihood estimation (MLE) with the method ``lstsq'' from the Qiskit Ignis API to reconstruct quantum states. The ``lstsq'' method first computes the least-squares estimate of the density matrix, then applies the
technique of \cite{smolin2012efficient} to impose positive semidefiniteness.

\begin{figure}[t]
\centering 
	\includegraphics[width=3.4in]{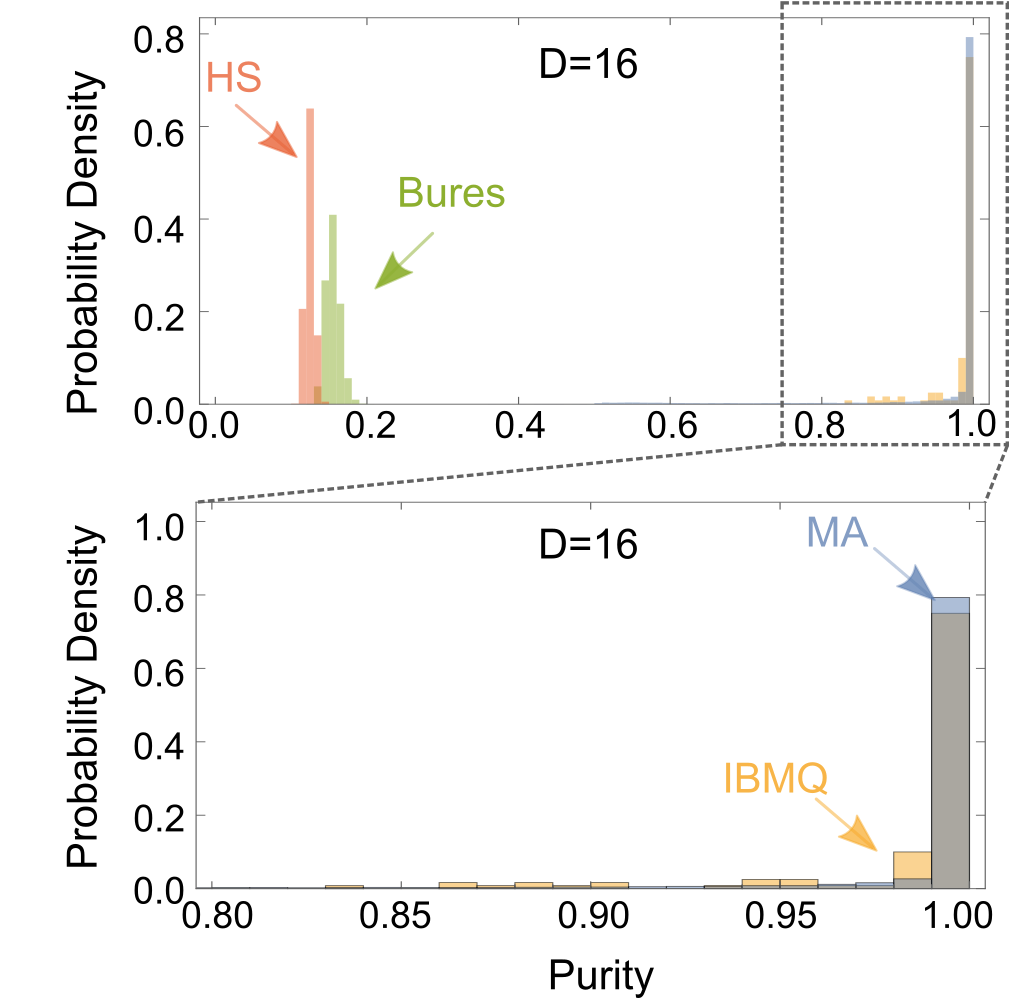}
\caption{Histograms comparing the Bures, HS, and MA distributions to the measured distribution of purity from the IBM Q for four-qubit quantum circuits initialized at $|0\rangle$ with no gates before tomography. The MA distribution has $D=K=16$ and $\alpha$ is tuned so that the mean of the distribution matches that of the measured IBM Q distribution.}
\label{Fig:ibmq-purity}
\end{figure}

\begin{figure}[!tb]
\centering 
	\includegraphics[width=3.4in]{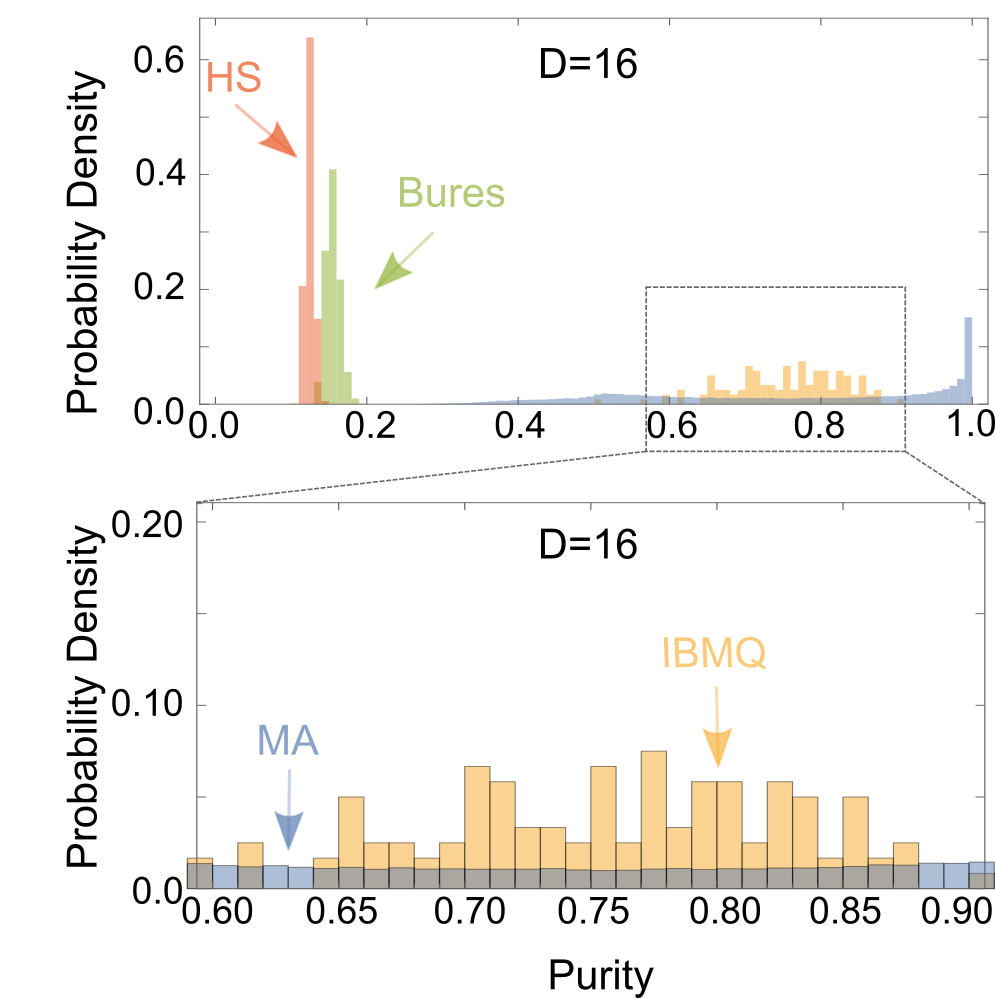}
\caption{Histograms comparing the Bures, HS, and MA distributions to the measured distribution of purity from the IBM Q for states resulting from random instances of the circuit shown in Fig. \ref{Fig:ibmq-intro}.  The MA distribution has $D=K=16$ and $\alpha$ is tuned so that the mean of the distribution matches that of the measured IBM Q distribution.}
\label{Fig:ibmq-purity2}
\end{figure}

Overall we perform complete state tomography 240 times (120 with no intermediate circuits, and 120 random circuits), 5000 shots each.
We then use these measurement results to reconstruct the corresponding density matrices and evaluate their purity. 
Histograms of the measured purity distributions found from the reconstructed states with no gates and with random instances of the circuit shown in Fig. \ref{Fig:ibmq-intro} are shown, respectively, in Figs. \ref{Fig:ibmq-purity} and \ref{Fig:ibmq-purity2}.
The mean purity of  IBM Q measured data in Figs. \ref{Fig:ibmq-purity} and \ref{Fig:ibmq-purity2} is 0.98 and 0.75, respectively.
For comparison, we also show the ($D=16$)-dimensional Bures and HS distributions, and the $D=K=16$ MA distribution with mean purity tuned to match the data sets obtained from IBM Q.
Each simulated histogram was created from $10^5$ random samples.

To quantitatively compare the distributions in Figs. \ref{Fig:ibmq-purity} and \ref{Fig:ibmq-purity2}, we also determine the overlap of the histograms in terms of the Bhattacharyya coefficients.
In particular, we find
\begin{equation}
    \begin{aligned}
    \mathcal{B}(\text{h}_{IBM1},\text{h}_{HS})&=\mathcal{B}(\text{h}_{IBM1},\text{h}_{B})=0\\
    \mathcal{B}(\text{h}_{IBM2},\text{h}_{HS})&=\mathcal{B}(\text{h}_{IBM2},\text{h}_{B})=0\\
    \mathcal{B}(\text{h}_{IBM1},\text{h}_{MA})&=0.92\\
    \mathcal{B}(\text{h}_{IBM2},\text{h}_{MA})&=0.55\\
    \end{aligned}
\end{equation}
In other words, after generating $10^5$ random states from both the Bures and HS distributions, neither resulted in a single state in the same bin as any of the measured states in either Figs. \ref{Fig:ibmq-purity} or \ref{Fig:ibmq-purity2}.
We also see a very high overlap between the MA distribution and the measured data in Fig. \ref{Fig:ibmq-purity}, where the average state has very high purity.

\subsection{Polarization-entangled photons}

\begin{figure}\centering 
	\includegraphics[width=.99\columnwidth]{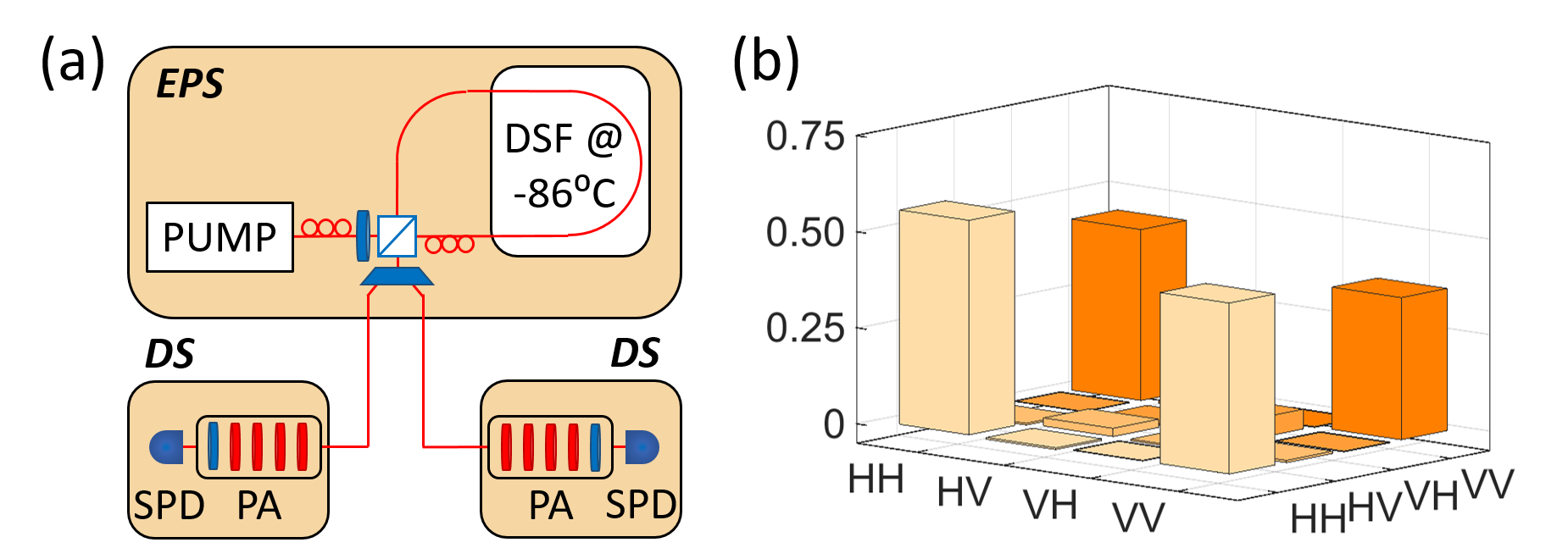}
	\caption{(a) Schematic of the setup used to characterize the distribution of quantum states generated by a polarization-entangled photon source (EPS). DSF: dispersion-shifted fiber. DS: detector station. PA: polarization analyzer consisting of several waveplates (red) and a polarizer (blue). SPD: single-photon detector. 
	(b) Mean density matrix resulting from performing quantum state tomography on the entangled state output by the EPS 1000 consecutive times.}
	\label{epsfig}
\end{figure}

A schematic diagram of the experimental setup used to characterize the distribution of quantum states generated by a polarization-entangled photon source is shown in Fig. \ref{epsfig}(a). The setup consists of an entangled-photon source (EPS) \cite{nucrypt} connected to two separate detector stations (DS) with telecom optical fibers. The EPS creates signal and idler photons via four-wave mixing \cite{fiorentino2002all} by pumping a dispersion-shifted fiber (DSF) with a 50 MHz pulsed fiber laser that operates at 1552.52 nm. The DSF is stored in a laboratory-grade freezer at $-86^{\circ}$C in order to reduce the generation of Raman-scattered noise photons. The average number of generated photon pairs per pulse can be tuned in the 0.001 – 0.1 range but is set at $\sim0.1$ here. By arranging the DSF in a Sagnac loop with a polarizing beam splitter (PBS), the signal and idler photons are entangled in polarization \cite{wang2009robust}. The signal and idler photons are then spectrally demultiplexed into 100 GHz-spaced ITU outputs, resulting in photons with a temporal duration of about 15 ps. For this experiment, we use channels 28 (1554.94 nm) and 34 (1550.12 nm).

\begin{figure}[!tb]
\centering 
	\includegraphics[width=3.4in]{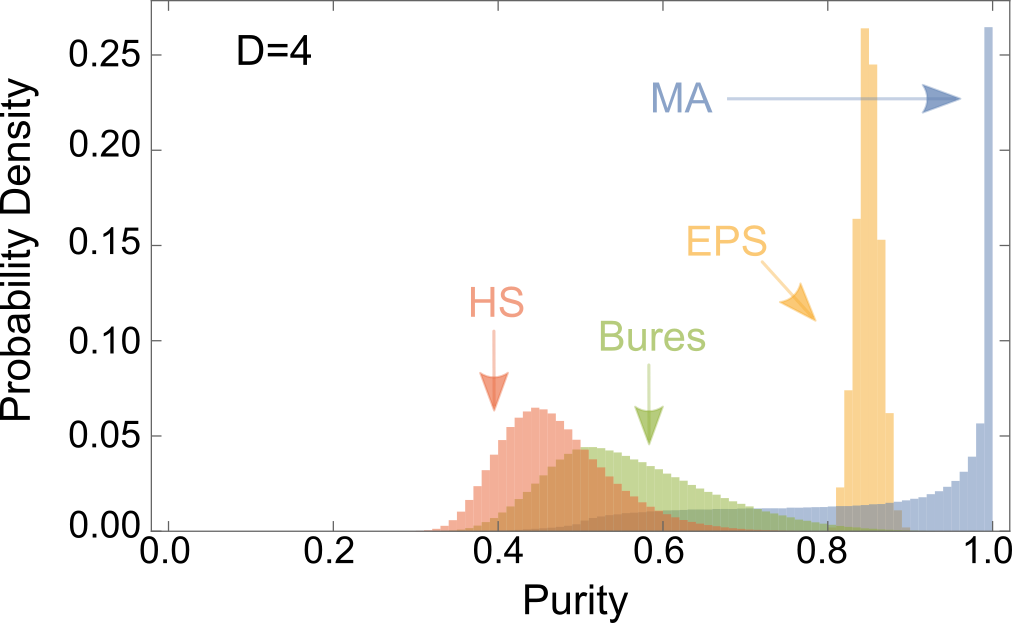}
\caption{Histograms comparing the Bures, HS, and MA distributions to the measured distribution of purity from a two-qubit entangled photon source (EPS).  The MA distribution has $D=K=4$ and $\alpha$ is tuned so that the mean of the distribution matches that of the measured EPS distribution.}
\label{Fig:EPS-purity}
\end{figure}

The detector stations include polarization analyzers (PA) and gated single-photon detectors (SPDs) with a detection efficiency of $\eta \sim 20\%$ and a dark count probability of $ \sim 4 \times 10^{-5}$ per gate. Automated FPGA-based software controls the detectors and analyzers in order to perform full polarization state tomography from measurements in $36$ different bases. Each of the 36 measurements is performed over 10 million detector gates, resulting in up to several thousand detected coincidences per measurement. The density matrix is then reconstructed using MLE~\cite{altepeter2005photonic}. The time required to perform all 36 measurements and MLE is about 12 s. Tomography is performed 1000 consecutive times, resulting in a total experiment duration of $\sim 4$ hr. The mean density matrix is shown in Fig. \ref{epsfig}(b). The purity is calculated for all 1000 density matrices.
Analogously to Figs. \ref{Fig:ibmq-purity} and \ref{Fig:ibmq-purity2}, we plot in Fig. \ref{Fig:EPS-purity} the histograms of the measured data, the $D=4$ HS and Bures distributions, and the $D=K=4$ MA distribution with mean purity tuned to match the EPS distribution.

To make a quantitative comparison between the various distributions in Fig. \ref{Fig:EPS-purity}, we have also calculated the Bhattacharyya coefficients to be

\begin{equation}
\begin{aligned}
    \mathcal{B}\left(\text{h}_{EPS},\text{h}_{HS}\right)&=0.006\\
    \mathcal{B}\left(\text{h}_{EPS},\text{h}_{B}\right)&=0.1\\
    \mathcal{B}\left(\text{h}_{EPS},\text{h}_{MA}\right)&=0.3\\
    \end{aligned}
\end{equation}
These results indicate that the MA distribution is closer to the measured EPS distribution than either the Bures or HS distances in terms of the Bhattacharyya coefficients.

\section{Discussion}

From a purely theoretical standpoint, the idea of performing inference with a prior (or training set) that does not correspond to the actual distribution of validation states seems unwarranted; after all, why should one intentionally select a prior that does not match the quantum states under investigation? And as we explored in Sec.~\ref{sec:exp}, the flexibility of the $MA$ distribution permits construction of priors or training sets that are well-tailored to an experimental system, so that such a situation of mismatched priors can be at least partially mitigated. Yet in some cases, such detailed prior specification may be undesirable.
For example, if one has strong beliefs that a quantum system produces pure maximally entangled states---after all, that was its design---it would be unwise to impose these beliefs on the prior in the context of state tomography: the goal is to \emph{show} that this belief is true from subsequent measurements, not \emph{assume} it \emph{a priori}. Accordingly, selection of an appropriate prior or training set might be motivated less by experimenter beliefs and more by the goal of providing a generic benchmark designed to let experiments guide the posterior distribution to the ground truth. 

In this practical sense, our examples highlight the value of exploring nontraditional priors and training sets for quantum state tomography. Even if lacking some of the theoretical properties of archetypal quantum state measures (like Bures), custom distributions like the MA distribution can attain performance comparable to these measures over the general Hilbert space---thus ensuring good fiducial uniformity as a test distribution---while enabling improvements for specific subspaces (e.g., pure states). Such performance hedging for desired outcomes bears resemblance to the recent estimation approach of classical shadows~\cite{huang2020predicting} which, when compared to BME, accepts much higher estimation error on average in exchange for remarkably low error for specific cases of interest~\cite{lukens2020bayesian}.

Interestingly, we have demonstrated measurable performance improvements by matching the MA distribution to only a single feature of the underlying distribution, the mean purity.  A topic of future research would be to determine how much further these results can be improved by matching the underlying distribution in more sophisticated ways, such as aiming to maximize overlap rather than merely matching means.  

The apparent recovery of the HS distribution by a properly tuned MA distribution offers, to our knowledge, a new and relatively simple physical interpretation of the HS distribution.  Traditionally, the HS distribution is motivated as naturally induced by tracing out Haar-random states of a higher dimension.  Our results indicate that alternatively, one can think of a $D$ dimensional HS distribution, resulting from a sum of $D$ Haar-random pure states with appropriate Dirichlet weights.  It would be interesting to determine if the distributions found via other induced measures can also be interpreted similarly by finding an appropriate $K$ and Dirichlet weights.  

\begin{acknowledgments}
A portion of this work was performed at Oak Ridge National Laboratory, operated by UT-Battelle for the U.S. Department of Energy under contract no. DE-AC05-00OR22725. J.M.L. acknowledges funding by the U.S. Department of Energy, Office of Science, Office of Advanced Scientific Computing Research, through the Quantum Algorithm Teams and Early Career Research Programs.  The views and conclusions contained in this document are those of the authors and should not be interpreted as representing the official policies, either expressed or implied, of the Army Research Laboratory or the U.S. Government. The U.S. Government is authorized to reproduce and distribute reprints for Government purposes notwithstanding any copyright notation herein. Additionally, we acknowledge use of the IBM Quantum for this work. The views expressed are those of the authors and do not reflect the official policy or position of IBM Quantum. This material is based upon work supported by, or in part by, the Army Research Laboratory and the Army Research Office under contract/grant numbers W911NF-19-2-0087 and W911NF-20-2-0168.   T.A.S. acknowledges support from the IBM-HBCU Quantum Center and the Martin Luther King Visiting Scholars Program at MIT.
\end{acknowledgments}

\appendix
\section*{Appendix}

We can create random density matrices according to \cite{zyczkowski1999volume} by taking $\bx$ of length $D$, creating a diagonal matrix from it, and rotating it with a Haar-random unitary from $U(D)$.
The purity of such a state is given by
\begin{equation}
    \text{Tr}\left(\rho^{2}\right)=\sum_{j=1}^{D}x_{j}^{2},
\end{equation}
and the expectation value of the purity by
\begin{equation}
    \text{E}_{Z}\left[\text{Tr}\left(\rho^{2}\right)
    \right]=\sum_{j=1}^{D}\text{E}\left[x_{j}^{2}\right]=\frac{1+\beta}{1+D\beta}
\end{equation}
where we have used $\beta$ as the concentration parameter of the Dirichlet distribution so as not to be confused with the MA expressions (where $\alpha$ was used).

\begin{figure}[!tb]
\centering 
	\includegraphics[width=3.4in]{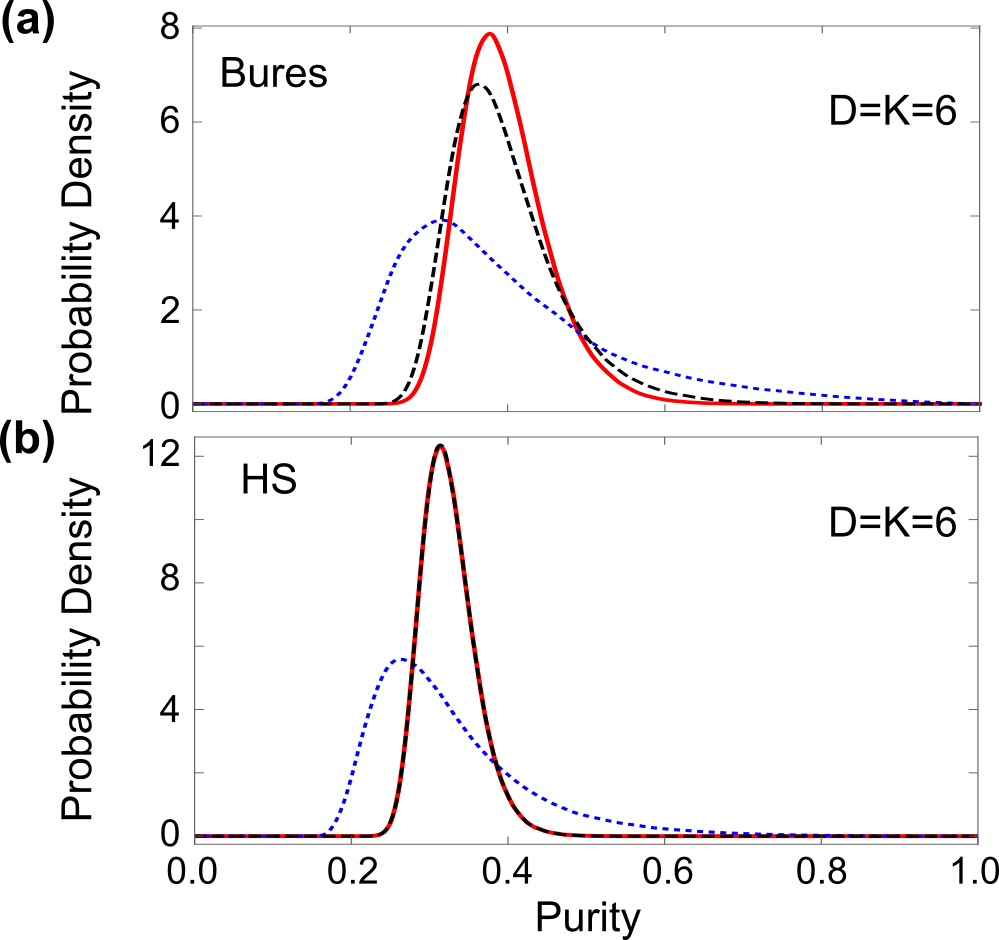}
\caption{Comparison of the Bures, HS, MA, and \.{Z}yczkowski distributions all with the same mean purity for $D=K=6$. Smoothed probability density functions are created from sampling each distribution $10^6$ times.  An analogous plot over several $D$ values is shown in Fig. \ref{Fig:comparison}. (a) The solid (red), dashed (black), and dotted (blue) lines correspond to the Bures, MA, and \.{Z}yczkowski distributions, respectively.  (b) The solid (red), dashed (black), and dotted (blue) lines correspond to the HS, MA, and \.{Z}yczkowski distributions, respectively.}
\label{fig:app}
\end{figure}

The $\beta$ values which tune the mean purity of this distribution to match that of the Bures, HS, and MA distributions [analogous to Eq. \eqref{eq:alpha_reproduce} for the MA distribution], are given by
\begin{equation}
    \begin{aligned}
    \beta_{MA}&=\frac{\alpha (D-1)}{(\alpha+1) D}\\
    \beta_{B}&=\frac{(D-1) (2 D-1)}{3 D (D+1)}\\
    \beta_{HS}&=\frac{D-1}{D+1}\\
    \end{aligned}
\end{equation}
In Fig. \ref{fig:app}, we overlay the numerically sampled and smoothed probability density functions of the Bures, HS, MA, and \.{Z}yczkowski distributions for $D=6$ ($K=D$ for MA). 
These plots are generated using $10^6$ random samples from each distribution, and the HS (solid red), Bures (solid Red), and MA (dashed black) lines are equivalent to the $D=6$ curves in Fig. \ref{Fig:comparison}.
In Fig. \ref{fig:app}(a), we compare the distribution of \.{Z}yczkowski (blue dotted) and MA (dashed black) distributions against the Bures distribution (solid red) when all have the same mean purity.  
Similarly, in Fig. \ref{fig:app}(b) we compare the distribution of \.{Z}yczkowski (blue dotted) and MA (dashed black) distributions against the HS distribution (solid red) when all have the same mean purity.  
We see that, unlike the MA distribution, the distribution of \.{Z}yczkowski does not reproduce either the HS or Bures distributions.
Further, the mode of \.{Z}yczkowski's distribution appears to be more mixed than any of the other three distributions for the same mean purity for the parameters plotted in Fig. \ref{fig:app}.

Finally, as was performed in Sec.~\ref{sec:exp}, we fit \.{Z}yczkowski's distribution by way of the mean purity to the data obtained in our experimental scenarios and compare with the MA distribution.
Adopting the same notation as in Sec.~\ref{sec:exp} we write the vector of histogram bins as $\text{h}_{s}$, where $s\in\{MA,B,HS,Z,IBM1,IBM2,EPS\}$ labels the distribution, with the new addition of $Z$ for \.{Z}yczkowski.
Calculating the Bhattacharyya coefficients we find
\begin{equation}
\begin{aligned}
    \frac{\mathcal{B}\left(\text{h}_{IBM1},\text{h}_{MA}\right)}{\mathcal{B}\left(\text{h}_{IBM1},\text{h}_{Z}\right)}&\approx \frac{\mathcal{B}\left(\text{h}_{IBM2},\text{h}_{MA}\right)}{\mathcal{B}\left(\text{h}_{IBM2},\text{h}_{Z}\right)}\approx 1\\
    \frac{\mathcal{B}\left(\text{h}_{EPS},\text{h}_{MA}\right)}{\mathcal{B}\left(\text{h}_{EPS},\text{h}_{Z}\right)}&=1.2\\
\end{aligned}
\end{equation}
In these expressions a value above one means that the MA distribution has a higher overlap with the experimental data, and a value of one means the two distributions are similar.
Hence, in all three cases we find that the MA distribution fits as well or slightly better than the distribution of \.{Z}yczkowski as measured by the Bhattacharyya coefficient when tuning to match mean purity.

\newpage
\bibliographystyle{ieeetr}
\bibliography{bibliography}

\begin{thebibliography}{10}

\bibitem{montanaro2007distinguishability}
A.~Montanaro, ``On the distinguishability of random quantum states,'' {\em
  Communications in Mathematical Physics}, vol.~273, no.~3, pp.~619--636, 2007.

\bibitem{hamma2012quantum}
A.~Hamma, S.~Santra, and P.~Zanardi, ``Quantum entanglement in random physical
  states,'' {\em Physical Review Letters}, vol.~109, no.~4, p.~040502, 2012.

\bibitem{miatto2015recovering}
F.~M. Miatto, K.~Pich{\'e}, T.~Brougham, and R.~W. Boyd, ``Recovering full
  coherence in a qubit by measuring half of its environment,'' {\em Physical
  Review A}, vol.~92, no.~6, p.~062331, 2015.

\bibitem{girolami2011quantum}
D.~Girolami and G.~Adesso, ``Quantum discord for general two-qubit states:
  analytical progress,'' {\em Physical Review A}, vol.~83, no.~5, p.~052108,
  2011.

\bibitem{kirby2016entanglement}
B.~T. Kirby, S.~Santra, V.~S. Malinovsky, and M.~Brodsky, ``Entanglement
  swapping of two arbitrarily degraded entangled states,'' {\em Physical Review
  A}, vol.~94, no.~1, p.~012336, 2016.

\bibitem{lu2011optimal}
X.-M. Lu, J.~Ma, Z.~Xi, and X.~Wang, ``Optimal measurements to access classical
  correlations of two-qubit states,'' {\em Physical Review A}, vol.~83, no.~1,
  p.~012327, 2011.

\bibitem{roncaglia2014bipartite}
M.~Roncaglia, A.~Montorsi, and M.~Genovese, ``Bipartite entanglement of quantum
  states in a pair basis,'' {\em Physical Review A}, vol.~90, no.~6, p.~062303,
  2014.

\bibitem{lu2018separability}
S.~Lu, S.~Huang, K.~Li, J.~Li, J.~Chen, D.~Lu, Z.~Ji, Y.~Shen, D.~Zhou, and
  B.~Zeng, ``Separability-entanglement classifier via machine learning,'' {\em
  Physical Review A}, vol.~98, no.~1, p.~012315, 2018.

\bibitem{lohani2020machine}
S.~Lohani, B.~T. Kirby, M.~Brodsky, O.~Danaci, and R.~T. Glasser, ``Machine
  learning assisted quantum state estimation,'' {\em Machine Learning: Science
  and Technology}, vol.~1, no.~3, p.~035007, 2020.

\bibitem{danaci2021machine}
O.~Danaci, S.~Lohani, B.~Kirby, and R.~T. Glasser, ``Machine learning pipeline
  for quantum state estimation with incomplete measurements,'' {\em Machine
  Learning: Science and Technology}, 2021.

\bibitem{lohani2020experimental}
S.~Lohani, T.~A. Searles, B.~T. Kirby, and R.~T. Glasser, ``On the experimental
  feasibility of quantum state reconstruction via machine learning,'' {\em
  arXiv:2012.09432}, 2020.

\bibitem{ahmed2020classification}
S.~Ahmed, C.~S. Mu{\~n}oz, F.~Nori, and A.~F. Kockum, ``Classification and
  reconstruction of optical quantum states with deep neural networks,'' {\em
  arXiv:2012.02185}, 2020.

\bibitem{Blume2010}
R.~Blume-Kohout, ``Optimal, reliable estimation of quantum states,'' {\em New
  Journal of Physics}, vol.~12, no.~4, p.~043034, 2010.

\bibitem{Seah2015}
Y.-L. Seah, J.~Shang, H.~K. Ng, D.~J. Nott, and B.-G. Englert, ``{Monte Carlo}
  sampling from the quantum state space. {II},'' {\em New Journal of Physics},
  vol.~17, p.~043018, apr 2015.

\bibitem{Granade2016}
C.~Granade, J.~Combes, and D.~G. Cory, ``Practical {Bayesian} tomography,''
  {\em New Journal of Physics}, vol.~18, no.~3, p.~033024, 2016.

\bibitem{Williams2017}
B.~P. Williams and P.~Lougovski, ``Quantum state estimation when qubits are
  lost: a no-data-left-behind approach,'' {\em New Journal of Physics},
  vol.~19, no.~4, p.~043003, 2017.

\bibitem{mai2017pseudo}
T.~T. Mai and P.~Alquier, ``Pseudo-{Bayesian} quantum tomography with
  rank-adaptation,'' {\em Journal of Statistical Planning and Inference},
  vol.~184, pp.~62--76, 2017.

\bibitem{lukens2020practical}
J.~M. Lukens, K.~J. Law, A.~Jasra, and P.~Lougovski, ``A practical and
  efficient approach for {Bayesian} quantum state estimation,'' {\em New
  Journal of Physics}, vol.~22, no.~6, p.~063038, 2020.

\bibitem{Lu2020b}
H.-H. Lu, E.~M. Simmerman, P.~Lougovski, A.~M. Weiner, and J.~M. Lukens,
  ``Fully arbitrary control of frequency-bin qubits,'' {\em Physical Review
  Letters}, vol.~125, p.~120503, Sep 2020.

\bibitem{wootters1990random}
W.~K. Wootters, ``Random quantum states,'' {\em Foundations of Physics},
  vol.~20, no.~11, pp.~1365--1378, 1990.

\bibitem{zyczkowski2001induced}
K.~Zyczkowski and H.-J. Sommers, ``Induced measures in the space of mixed
  quantum states,'' {\em Journal of Physics A: Mathematical and General},
  vol.~34, no.~35, p.~7111, 2001.

\bibitem{Sommers2003}
H.-J. Sommers and K.~\.{Z}yczkowski, ``Bures volume of the set of mixed quantum
  states,'' {\em Journal of Physics A: Mathematical and General}, vol.~36,
  pp.~10083--10100, sep 2003.

\bibitem{zyczkowski2005average}
K.~{\.Z}yczkowski and H.-J. Sommers, ``Average fidelity between random quantum
  states,'' {\em Physical Review A}, vol.~71, no.~3, p.~032313, 2005.

\bibitem{al2010random}
V.~Al~Osipov, H.-J. Sommers, and K.~{\.Z}yczkowski, ``Random {Bures} mixed
  states and the distribution of their purity,'' {\em Journal of Physics A:
  Mathematical and Theoretical}, vol.~43, no.~5, p.~055302, 2010.

\bibitem{alquier2015bayesian}
T.~T. Mai and P.~Alquier, ``A {Bayesian} approach for noisy matrix completion:
  Optimal rate under general sampling distribution,'' {\em Electronic Journal
  of Statistics}, vol.~9, no.~1, pp.~823--841, 2015.

\bibitem{cottet20181}
V.~Cottet and P.~Alquier, ``1-bit matrix completion: {PAC-Bayesian} analysis of
  a variational approximation,'' {\em Machine Learning}, vol.~107, no.~3,
  pp.~579--603, 2018.

\bibitem{Lingaraju2021}
N.~B. Lingaraju, H.-H. Lu, S.~Seshadri, D.~E. Leaird, A.~M. Weiner, and J.~M.
  Lukens, ``Adaptive bandwidth management for entanglement distribution in
  quantum networks,'' {\em Optica}, vol.~8, pp.~329--332, Mar 2021.

\bibitem{Alshowkan2021}
M.~Alshowkan, B.~P. Williams, P.~G. Evans, N.~S. Rao, E.~M. Simmerman, H.-H.
  Lu, N.~B. Lingaraju, A.~M. Weiner, C.~E. Marvinney, Y.-Y. Pai, B.~J. Lawrie,
  N.~A. Peters, and J.~M. Lukens, ``A reconfigurable quantum local area network
  over deployed fiber,'' {\em arXiv:2102.13596}, 2021.

\bibitem{ekert2002direct}
A.~K. Ekert, C.~M. Alves, D.~K. Oi, M.~Horodecki, P.~Horodecki, and L.~C. Kwek,
  ``Direct estimations of linear and nonlinear functionals of a quantum
  state,'' {\em Physical Review Letters}, vol.~88, no.~21, p.~217901, 2002.

\bibitem{zyczkowski1999volume}
K.~{\.Z}yczkowski, ``Volume of the set of separable states. ii,'' {\em Physical
  Review A}, vol.~60, no.~5, p.~3496, 1999.

\bibitem{fuchs1999cryptographic}
C.~A. Fuchs and J.~Van De~Graaf, ``Cryptographic distinguishability measures
  for quantum-mechanical states,'' {\em IEEE Transactions on Information
  Theory}, vol.~45, no.~4, pp.~1216--1227, 1999.

\bibitem{bengtsson2017geometry}
I.~Bengtsson and K.~{\.Z}yczkowski, {\em Geometry of Quantum States: An
  Introduction to Quantum Entanglement}.
\newblock Cambridge University Press, 2017.

\bibitem{alonso2012ehrenfest}
J.~L. Alonso, J.~Clemente-Gallardo, J.~C. Cuch{\'\i}, P.~Echenique, and
  F.~Falceto, ``Ehrenfest dynamics is purity non-preserving: A necessary
  ingredient for decoherence,'' {\em The Journal of Chemical Physics},
  vol.~137, no.~5, p.~054106, 2012.

\bibitem{Telgarsky2013}
M.~Telgarsky, ``Dirichlet draws are sparse with high probability,'' {\em
  arXiv:1301.4917}, 2013.

\bibitem{bhatia2000better}
R.~Bhatia and C.~Davis, ``A better bound on the variance,'' {\em The American
  Mathematical Monthly}, vol.~107, no.~4, pp.~353--357, 2000.

\bibitem{ginibre1965statistical}
J.~Ginibre, ``Statistical ensembles of complex, quaternion, and real
  matrices,'' {\em Journal of Mathematical Physics}, vol.~6, no.~3,
  pp.~440--449, 1965.

\bibitem{sommers2004statistical}
H.-J. Sommers and K.~{\.Z}yczkowski, ``Statistical properties of random density
  matrices,'' {\em Journal of Physics A: Mathematical and General}, vol.~37,
  no.~35, p.~8457, 2004.

\bibitem{melnikov2018active}
A.~A. Melnikov, H.~P. Nautrup, M.~Krenn, V.~Dunjko, M.~Tiersch, A.~Zeilinger,
  and H.~J. Briegel, ``Active learning machine learns to create new quantum
  experiments,'' {\em Proceedings of the National Academy of Sciences},
  vol.~115, no.~6, pp.~1221--1226, 2018.

\bibitem{sentis2015quantum}
G.~Sent{\'\i}s, M.~Gu{\c{t}}{\u{a}}, and G.~Adesso, ``Quantum learning of
  coherent states,'' {\em EPJ Quantum Technology}, vol.~2, no.~1, pp.~1--22,
  2015.

\bibitem{harney2020entanglement}
C.~Harney, S.~Pirandola, A.~Ferraro, and M.~Paternostro, ``Entanglement
  classification via neural network quantum states,'' {\em New Journal of
  Physics}, vol.~22, no.~4, p.~045001, 2020.

\bibitem{bharti2020machine}
K.~Bharti, T.~Haug, V.~Vedral, and L.-C. Kwek, ``Machine learning meets quantum
  foundations: A brief survey,'' {\em AVS Quantum Science}, vol.~2, no.~3,
  p.~034101, 2020.

\bibitem{MacKay2003}
D.~J.~C. MacKay, {\em {Information Theory, Inference, and Learning
  Algorithms}}.
\newblock Cambridge, UK: Cambridge University Press, 2003.

\bibitem{Robert1999}
C.~P. Robert and G.~Casella, {\em Monte Carlo Statistical Methods}.
\newblock New York: Springer, 1999.

\bibitem{Mezzadri2007}
F.~Mezzadri, ``How to generate random matrices from the classical compact
  groups,'' {\em Notices of the American Mathematical Society}, vol.~54,
  p.~592, 2007.

\bibitem{Cotter2013}
S.~L. Cotter, G.~O. Roberts, A.~M. Stuart, and D.~White, ``{MCMC} methods for
  functions: Modifying old algorithms to make them faster,'' {\em Statistical
  Science}, vol.~28, pp.~424--446, 08 2013.

\bibitem{Lewis1986}
P.~A. Lewis, E.~McKenzie, and D.~K. Hugus, ``Gamma processes,'' tech. rep.,
  Naval Postgraduate School, 1986.

\bibitem{Qiskit}
H.~Abraham, AduOffei, R.~Agarwal, I.~Y. Akhalwaya, G.~Aleksandrowicz,
  T.~Alexander, M.~Amy, E.~Arbel, Arijit02, A.~Asfaw, A.~Avkhadiev,
  C.~Azaustre, AzizNgoueya, A.~Banerjee, A.~Bansal, P.~Barkoutsos, A.~Barnawal,
  G.~Barron, G.~S.~B. ..., and M.~{\v{C}}epulkovskis, ``Qiskit: An open-source
  framework for quantum computing,'' 2019.

\bibitem{smolin2012efficient}
J.~A. Smolin, J.~M. Gambetta, and G.~Smith, ``Efficient method for computing
  the maximum-likelihood quantum state from measurements with additive gaussian
  noise,'' {\em Physical Review Letters}, vol.~108, no.~7, p.~070502, 2012.

\bibitem{nucrypt}
NuCrypt, ``Quantum optical instrumentation.''

\bibitem{fiorentino2002all}
M.~Fiorentino, P.~L. Voss, J.~E. Sharping, and P.~Kumar, ``All-fiber
  photon-pair source for quantum communications,'' {\em IEEE Photonics
  Technology Letters}, vol.~14, no.~7, pp.~983--985, 2002.

\bibitem{wang2009robust}
S.~X. Wang and G.~S. Kanter, ``Robust multiwavelength all-fiber source of
  polarization-entangled photons with built-in analyzer alignment signal,''
  {\em IEEE Journal of Selected Topics in Quantum Electronics}, vol.~15, no.~6,
  pp.~1733--1740, 2009.

\bibitem{altepeter2005photonic}
J.~B. Altepeter, E.~R. Jeffrey, and P.~G. Kwiat, ``Photonic state tomography,''
  {\em Advances in Atomic, Molecular, and Optical Physics}, vol.~52,
  pp.~105--159, 2005.

\bibitem{huang2020predicting}
H.-Y. Huang, R.~Kueng, and J.~Preskill, ``Predicting many properties of a
  quantum system from very few measurements,'' {\em Nature Physics}, vol.~16,
  no.~10, pp.~1050--1057, 2020.

\bibitem{lukens2020bayesian}
J.~M. Lukens, K.~J. Law, and R.~S. Bennink, ``A {Bayesian} analysis of
  classical shadows,'' {\em arXiv:2012.08997}, 2020.

\end{thebibliography}

\end{document}